\documentclass[useAMS, usenatbib, usegraphicx]{mn2e}
\title[X-ray -- age relation and exoplanet evaporation]{The coronal X-ray -- age relation and its implications for the evaporation of exoplanets}
\author[A.P. Jackson, T.A. Davis and P.J. Wheatley]{Alan P. Jackson$^{1,2}$\thanks{E-mail: ajackson@ast.cam.ac.uk (APJ); tdavis@eso.org (TAD)}, Timothy A. Davis$^{1,3}$\footnotemark[1] and Peter J. Wheatley$^4$\\%
$^1$Sub-Department of Astrophysics, University of Oxford, Denys Wilkinson Building, Keble Road, Oxford, UK, OX1 3RH\\%
$^2$Institute of Astronomy, University of Cambridge, Madingley Road, Cambridge, UK, CB3 0HA\\%
$^3$European Southern Observatory, Karl-Schwarzschild-Str. 2, 85748, Garching bei Muenchen, Germany\\%
$^4$Department of Physics, University of Warwick, Gibbet Hill Road, Coventry, UK, CV4 7AL}
\date{Submitted 2011}
\pagerange{\pageref{firstpage}--\pageref{lastpage}}
\pubyear{2011}
\usepackage[total={17.8cm,24.0cm},centering]{geometry}
\usepackage{times}
\usepackage[fleqn]{amsmath}
\usepackage[none]{hyphenat}

\begin{document}
\label{firstpage}
\maketitle
\begin{abstract}
We study the relationship between coronal X-ray emission and stellar age for late-type stars, and the variation of this relationship with spectral type.  We select 717 stars from 13 open clusters and find that the ratio of X-ray to bolometric luminosity during the saturated phase of coronal emission decreases from $10^{-3.1}$ for late K-dwarfs to $10^{-4.3}$ for early F-type stars (across the range \mbox{$0.29\leq (B-V)_0<1.41$}).  Our determined saturation timescales vary between $10^{7.8}$ and $10^{8.3}$ years, though with no clear trend across the whole FGK range.\\
We apply our X-ray emission -- age relations to the investigation of the evaporation history of 121 known transiting exoplanets using a simple energy-limited model of evaporation and taking into consideration Roche lobe effects and different heating/evaporation efficiencies.  We confirm that a linear cut-off of the planet distribution in the $M^2/R^3$ versus $a^{-2}$ plane is an expected result of population modification by evaporation and show that the known transiting exoplanets display such a cut-off.  We find that for an evaporation efficiency of 25 per cent we expect around one in ten of the known transiting exoplanets to have lost $\geq5$ per cent of their mass since formation.  In addition we provide estimates of the minimum formation mass for which a planet could be expected to survive for 4~Gyrs for a range of stellar and planetary parameters.\\
We emphasise the importance of the earliest periods of a planet's life for it's evaporation history with 75 per cent expected to occur within the first Gyr.  This raises the possibility of using evaporation histories to distinguish different migration mechanisms.  For planets with spin-orbit angles available from measurements of the Rossiter-McLaughlin effect no difference is found between the distributions of planets with misaligned orbits and those with aligned orbits.  This suggests that dynamical effects accounting for misalignment occur early in the life of the planetary system, although additional data is required to test this.
\end{abstract}

\section{Introduction}
\label{introduction}
Since the discovery of the first exoplanet (51 Peg.\ b) around a main sequence star by \citet{mayor}, a roughly Jupiter mass planet in a $\sim$4.2 day orbit, many similar 'hot Jupiter' type exoplanets have been discovered.  This has provoked much work to investigate where and how these hot Jupiters form and how they might migrate to their current positions (see \citealt{papaloizou} for a review).

Intriguingly some exoplanetary properties appear to be correlated; possible correlations between planet mass and surface gravity with orbital period have been reported by \citet*{mazeh} and \citet*{southworth} respectively.  Both correlations have subsequently been re-plotted by a number of authors as more exoplanets have been discovered (e.g.~\citealt{hansen, pollacco, davis}) and are reproduced in Section \ref{planetsample} for the sample of exoplanets used in this work.

One possible mechanism for producing such correlations is the evaporation of close orbiting planets, such as that observed for HD209458b (also known as Osiris, e.g.~\citealt{vidalmadjar2}) by \citet{vidalmadjar1}\footnote{see also \citet{benjaffel, vidalmadjar2}} and HD189733b by \citet{lecavelier2}.

A simple model of exoplanet evaporation was developed by \citet{lecavelier1}, driven by X-ray/Extreme Ultraviolet (EUV) radiation.  He applied it to the known exoplanets and found that they should not be losing significant mass at today's irradiation levels.  The X-ray/EUV irradiation level will be much higher around a younger solar-type star however, as discussed further below, and \citet*{penz} showed that, in principle, it is possible for evaporation to significantly effect the mass distribution of close orbiting exoplanets.

As already noted by \citet{lecavelier1} and \citet{penz} it is clear that to understand the evaporation of close orbiting exoplanets an understanding of the X-ray/EUV emission of the host star and its evolution is essential.  \citet{davis}\defcitealias{davis}{DW09} and others such as \citet{lammer2009}\defcitealias{lammer2009}{L09} (hereafter \citetalias{davis} and \citetalias{lammer2009} respectively) combined the energy-limited model developed by \citet{lecavelier1} with simple, coarse, estimates of the X-ray evolution to study the possible evaporation histories of known transiting exoplanets.

The majority of the X-ray/EUV luminosity of a star is emitted from the hot corona (with typical temperatures of order 1 keV), the heating of which is believed to be related to magnetic activity within the star (e.g. \citealt{erdelyi2007}).  The stellar magnetic fields, of which this activity and X-ray emission is a manifestation, are themselves thought to be derived from complex dynamo mechanisms at work in the interior of the star.  The efficiency of these mechanisms is determined by the interaction between convection in the outer envelope and differential rotation (e.g.~\citealt*{landstreet, covas, donati}).  \citet{skumanich} observed a proportionality between average surface magnetic field and stellar rotation rate and was the first to suggest a relationship between activity and rotation as a consequence of the dynamo mechanism.  Since then many authors, such as \citet{pallavicini1981}, \citet{randich3} and \citet{pizzolato} (hereafter\defcitealias{pizzolato}{P03}\citetalias{pizzolato}), have studied the relationship between activity/X-ray emission and rotation.

Over time the rotation rate of a star slows as a result of magnetic braking (e.g. \citealt{ivanova}) and since the magnetic dynamo is linked to the stellar rotation the magnetic field will also decrease over time.  Thus it would be expected that the X-ray luminosity will fall with decreasing rotation rate and indeed this is observed, e.g.~\citetalias{pizzolato} see a decrease for periods longer than $\sim$1-2 days.  This trend of increasing X-ray luminosity/magnetic activity with increasing rotation rate cuts off at short rotation periods ($\sim$1-2 days).  \citetalias{pizzolato}, and others (e.g.~\citealt{micela}), find that at rotation periods shorter than this turn-off point X-ray emission ceases to increase with rotation rate and instead becomes saturated.  Since the rotation rate decreases with time this means that there will also be an age at which stars will turn off the saturated regime and after which the X-ray luminosity will begin to fall off.

In this work we aim to improve constraints on the direct relationship between stellar X-ray luminosity and age for late-type stars.  We then apply this to the study of the evaporation of exoplanets to extend on previous work such as that of \citetalias{davis} and \citetalias{lammer2009} which used rather coarse estimates of the X-ray evolution.  In Section \ref{Xray} we present our study of the evolution of the X-ray emission of late-type stars with the investigation of exoplanet evaporation following in Section \ref{exoplanets}.  Finally in Section \ref{conclusion} we present our conclusions.

\section{X-ray evolution of late-type stars}
\label{Xray}
\subsection{Cluster/star sample}
\label{starsample}
To obtain constraints on the evolution of X-ray emission from solar type stars we require stars of known age across a range of ages.  Therefore we have selected stars from open clusters with age estimates, the clusters being selected such that they cover a range of ages from $\sim5-740$ Myr.  We utilise open clusters since they form in a short space of time, and thus all of the members will be of similar age.  Details of the clusters can be found in Table \ref{clusters} and a complete catalogue of the stars used in this work is available online through the VizieR database\footnote{VizieR data web address}.  The star/cluster data is taken from past X-ray surveys of open clusters in the literature.  Most of the original observations were taken using the PSPC or HRI instruments on \emph{ROSAT (R\"ontgen Satellit)}, though some of the more recent surveys were conducted using the \emph{Chandra} and \emph{XMM-Newton} telescopes.

\begin{table*}
\begin{minipage}{150mm}
\caption{Details of the open clusters used in this work and the source of the stellar data for each.}
\label{clusters}
\begin{tabular}{c r@{}l r@{}l r@{}l l r c}
\hline
Cluster &\multicolumn{2}{c}{Age (Myr)}& \multicolumn{2}{c}{Distance (pc)}&\multicolumn{2}{c}{Apparent distance}	&$E(B-V)$	  & No. of & Stellar data source\\
	& &                           & &                               &\multicolumn{2}{c}{modulus (mag)}	&\hspace{8pt}(mag)& stars & \\
\hline
$\alpha$ Persei & 79&$_{a,b,c}$ &	182&$_{a,b}$ 		&	6&.587$_{a,b}$ 		&	0.090$_{a,b}$ &	66&	\citet{randich3}, \citetalias{pizzolato}\\
Blanco 1 &	100&$_{a,d}$    &	244&$_{a,b}$ 		&	7&.18$_{a}$ 		&	0.01$_{a}$ &	29&	\citet{cargile}\\
Hyades &	630&$_{b,e}$    &	46&.3$_{e}$ 		&	3&.42$_{b}$ 		&	0.01$_{b}$ &	82&	\citet*{stern}\\
IC2391 &	50&$_{a,b,c}$   &	179&$_{a,b}$ 		&	6&.30$_{a,b}$ 		&	0.01$_{a,b}$ &	19&	\citet{patten}, \citetalias{pizzolato}\\
IC2602 &	37&$_{a,b,f}$	&	163&$_{a,b}$		&	6&.145$_{a,b}$ 		&	0.026$_{a,b}$ &	26&	\citet{randich2}, \citetalias{pizzolato}\\
NGC1039 &	192&$_{a,b}$	&	491&$_{a,b}$ 		&	8&.681$_{a,b}$ 		&	0.07$_{a,b}$ &	9&	\citet{simon}\\
NGC2516 &	114&$_{a,b}$	&	361&$_{a,b}$ 		&	8&.062$_{a,b}$ 		&	0.086$_{a,b}$ &	97&	\citet{damiani1}\\
NGC3532 &	308&$_{a,b}$	&	496&$_{a,b}$ 		&	8&.599$_{a,b}$ 		&	0.039$_{a,b}$ & 9&	\citet{franciosini}\\
NGC2547 &	34&$_{a,b,g,h}$	& 	420&$_{a,b,g,h}$ 	& 	8&.41$_{a,b}$ 		& 	0.039$_{a,b}$ & 41&	\citet{jeffries2}\\
NGC6475 &	197&$_{a,b}$	&	283&$_{a,b}$ 		&	7&.584$_{a,b}$ 		&	0.102$_{a,b}$ &	59&	\citet{prosser}\\
NGC6530 &	5&.4$_{a,b,i}$	&	1062&$_{a,b}$ 		&	11&.138$_{a,b}$ 	&	0.315$_{a,b}$ &	171&	\citet{damiani2}\\
Pleiades &	132&$_{a,b}$	&	134&$_{k}$ 		&	-&- 			&	0.025$_{a,b}$ &	95&	\citet{stauffer1}\\
Praesepe &	741&$_{a,b}$	&	194&$_{a,b}$ 		&	6&.468$_{a,b}$ 		&	0.01$_{a,b}$ &	14&	\citet{randich1}\\
\hline
\end{tabular}\\
\emph{References}: a) \citet{kharchenko}, b) \citet*{loktin}, c) \citet*{barradoy}, d) \citet*{cargile}, James et al. (2009, in prep.), e) \citet{perryman}, f) \citet{stauffer2}, g) \citet{naylor}, h) \citet{jeffries1},   i) \citet{damiani2}, k) \citet{soderblom2}.\\*
Where more than one reference is given the value shown is a weighted mean of the reference values.  Where the distance, apparent distance modulus and $E(B-V)$ references are the same the distance is calculated from the apparent distance modulus and $E(B-V)$.
\end{minipage}
\end{table*}

Within the datasets for each cluster we select stars that fall in the range \mbox{$0.29\leq (B-V)_0<1.41$} (roughly stars of spectral type F, G and K), which have well defined X-ray luminosities (i.e.\ not upper limits).  We exclude stars with only upper limits since these are presented inhomogeneously or not at all in the sources from which we draw our sample, and a meaningful comparison between these estimates could not be made.  Where cluster membership information is given we require the stars to be likely cluster members/have a membership probability of $>70\%$ for selection.

In addition to the data from the open clusters listed in Table \ref{clusters} we also include in our plots the field star sample of \citetalias{pizzolato} as a guide to how our results relate to the properties of stars of Solar age.  We do not use these stars in any of the fitting however.  A list of the stars in this field star sample can be found in Table 1 of \citetalias{pizzolato}

\subsubsection{Determination of X-ray and bolometric luminosities}
\label{xraybollum}
Where X-ray luminosities have been calculated in the source papers, as they are for all of the clusters except NGC 6530, we use the luminosities given.  For NGC 6530 we use the web\textsc{pimms} tool available at NASA HEASARC\footnote{heasarc.gsfc.nasa.gov} to determine the count rate to flux conversion factor.  The column density of hydrogen, $N_H$, is obtained from the cluster $E(B-V)$ using the formula $N_H=[E(B-V)\times 5.8\times 10^{21}]$ cm$^{-2}$ (\citealt{bohlin1978}).  Once we have the cluster distance the X-ray luminosities can then be determined.

With the exception of the Pleiades and Hyades (which have parallax distances) we calculate the cluster distance from the apparent distance modulus in conjunction with the reddening, $E(B-V)$, and the extinction law $A_V=3.1\times E(B-V)$ (e.g. \citealt{schultz1975, sneden1978}).  Where these distance estimates (from apparent distance modulus or parallax) differ from those used in the original study, and are based on more recent data, we correct the X-ray luminosities for the new distance.

Spectral type information is not available for the majority of stars used in this work so we use the de-reddened colour indices, $(B-V)_0=(B-V)-E(B-V)$, to assign spectral types using the tables presented in \citet{lang}.  Where necessary we calculate bolometric luminosities using the absolute magnitude of the star.  We determine the absolute magnitude from the apparent $V$ magnitude and the cluster apparent distance modulus in combination with a bolometric correction calculated using the de-reddened colour index, $(B-V)_0$, and the tables presented in \citet{lang}.  The luminosities are calibrated using an absolute bolometric magnitude of +4.75 for the Sun, as given in \citet{lang}.  We thus calculate the X-ray to bolometric luminosity ratio where this has not been done in the source paper.

\subsubsection{Errors in X-ray and bolometric luminosities}
\label{stellarXerrors}
Within each $(B-V)_0$ bin for each cluster there will be (provided that there is more than one star present in that bin) a scatter in the X-ray to bolometric luminosity ratios.  There will also be errors in the luminosity ratios of the individual stars and these may be similar in magnitude to the intrinsic scatter.

Where errors in the stellar X-ray luminosities are given in the source papers we use them as provided, where errors are only given in the count rates or X-ray luminosities are not calculated in the source papers we estimate the error by propagating through the count rate and distance errors.  As $E(B-V)$ and $N_H$ are correction factors, and are for our sample generally quite small with fairly small fractional errors, we neglect errors in these quantities in our error analysis.  The X-ray luminosity errors are thus determined solely by the count rate and distance errors.

In the case of bolometric luminosities there will be a contribution from the error in the cluster apparent distance modulus and also contributions from errors in the apparent $V$ magnitude and $B-V$ colour (as before we neglect errors in $E(B-V)$).  Unfortunately however the only cluster for which we have data on errors in $V$ magnitude and $B-V$ is Blanco 1.  As such, and with the Blanco 1 errors as a rough guide, we decide the most reasonable course is to estimate the error in the $V$ magnitude and $B-V$ colours of the other clusters as being $\pm 1$ in the last digit.  The estimated $B-V$ colour error gives an estimate of the error in the bolometric correction.  The error in the bolometric luminosity will thus be determined by the error in the cluster apparent distance modulus and the estimated $V$ magnitude and bolometric correction errors.

Having calculated the errors in the X-ray and bolometric luminosities we can determine an error in the luminosity ratio for each star and then calculate the mean stellar error within each $(B-V)_0$ bin for each cluster.  We then combine this with the standard deviation of the stars from the mean luminosity ratio in the relevant $(B-V)_0$ bin of the cluster to determine a total luminosity ratio error within each $(B-V)_0$ bin for each cluster.  We note that for the most well populated clusters this may lead to a slight over-estimation of the errors since some of the scatter between different stars in the cluster will be due to the errors in the determination of X-ray and bolometric luminosities for the individual stars.  For many of the clusters in our sample however there are not enough stars for the cluster error to fully account for the errors in individual stars.  To ensure that all of the clusters are treated equally we thus combine the cluster and individual star errors for all clusters.  Any over-estimation of the errors for the most well populated clusters should be accounted for by the weighting described in Section \ref{regimefits}.

\subsubsection{Stars excluded from the data}
\label{exclusioncriteria}
As the X-ray surveys are complete surveys of the entire clusters they do not make any distinctions regarding main-sequence (MS) stars, pre-main sequence (PMS) stars, post-main sequence giants or variables.  We consider that the processes leading to X-ray emission in PMS stars are likely to be similar to MS stars and thus do not exclude them, indeed in our younger clusters the majority of stars will be PMS stars.  Nonetheless we do take note of different classes of PMS stars, in particular variables, and discuss the potential effects of these stars on the mean X-ray characteristics in Section \ref{binandvar}.

On the other hand we would not expect the dynamo processes that drive coronal X-ray emission to be the same in giants as in MS stars and thus would not expect them to display the same X-ray properties.  As such giant stars should be excluded from our sample.  Giant stars are also comparatively easy to identify even where the clusters have not been well studied at other wavelengths due to their greater luminosity.  We thus exclude any star which would have a bolometric luminosity $>$100 times that of the Sun if it is at the associated cluster distance.  As well as giants this will also help to filter out foreground contaminants of the cluster sample.  For variable stars we want to exclude contact/interacting binary systems, such as W UMa type systems, since it is likely that the interaction between the stars will have significant effects on the X-ray characteristics.

Given the size of the cluster datasets, and that some of our clusters have not been well studied as yet, it was not possible to perform checks on all of the stars in each cluster to indentify stars falling into one of the above exclusion criteria.  As such we chose to check the Pleiades, Praesepe and Hyades datasets as these clusters are the most well studied at other wavelengths in our sample and use these to inform subsequent targeted checks on the other datasets.  The checking process consisted of taking the star designation/coordinates from the survey data, querying the SIMBAD\footnote{simbad.u-strasbg.fr/simbad; see also \citet{wenger}} database and analysing the details returned.  From this checking process we found that in general stars falling into an exclusion category are outliers, and did not show any preferential distribution when mixed in with the main bulk of the clusters.  As such for the remaining clusters we focused checks on stars which appear to be outliers.  As already mentioned however some of the other clusters in our sample have not yet been well studied and so lack the detailed information on member stars necessary to identify exclusion candidates.  None of our exclusion criteria result in the exclusion of a very large number of stars though ($< 10\%$ from the Pleiades, Praesepe and Hyades in total for all exclusion criteria) so the lack of information to identify exclusion candidates in some of the clusters should simply increase the scatter in the affected clusters and not alter the mean values significantly.

In addition to the above general categories we also exclude stars that are identified as flaring during the observations, since by definition a flare will increase the X-ray luminosity of the star significantly above the quiescent level.  Significant contamination of the sample by flares would thus cause the mean X-ray luminosities to be over-estimated as well as increasing the scatter in our relations.  Where available we use studies of flare stars to exclude those stars that were in flare during the observations.  For the Pleiades we use the study of \citet*{gagne}, for NGC2516 we use \citet{wolk} and for NGC2547 we use \citet{jeffries2}.  We do not exclude 'flare stars' as a general class however and discuss the effect this may have on the mean X-ray characteristics in Section \ref{binandvar}.  \citet{gagne} and \citet{jeffries2} estimate a flaring rate of approximately one flare every 350 ks for flare stars of ages between that of NGC2547 and the Pleiades.  We thus expect that contamination by flares will not be too great for clusters at these ages or older since flaring is more strongly associated with PMS stars, and we expect all of the stars in these clusters to have joined the MS\@.  \citet{damiani2} identify a higher flare rate in NGC6530, the youngest cluster, and also identify the highest flare rates with the youngest stars, though they do not describe individual flares in detail.  As well as utilising these studies we exclude any star found to have an X-ray to bolometric luminosity ratio of $>10^{-2.5}$ as highly likely to be in flare, as this is well above the saturated level found by any author.

\subsection{Analysis of the X-ray evolution}
\label{Xray analysis}
We expect that the behaviour of the X-ray luminosity will change with stellar mass and with bolometric luminosity.  Thus for our analysis of the X-ray evolution we divide our data into bins by $(B-V)_0$ colour as a measurable, continuous, proxy for these properties.

\begin{figure}
\includegraphics[width=85mm]{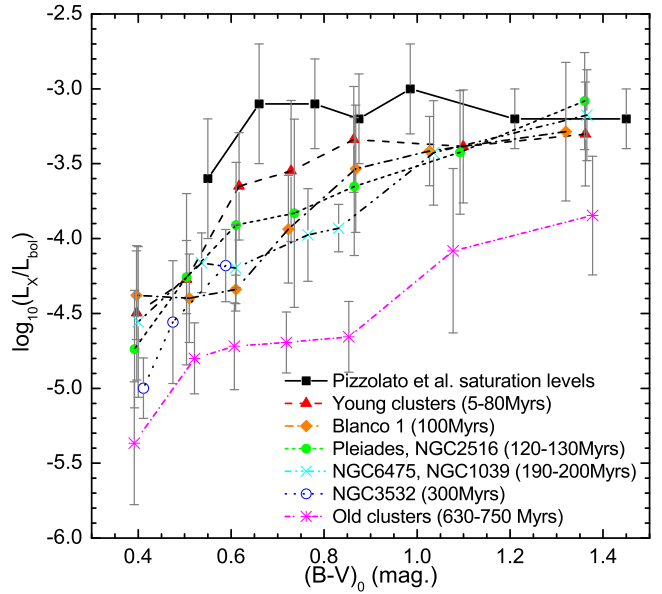}
\caption{$\log (L_X/L_{bol})$ against $(B-V)_0$ colour for all clusters in this work and the saturation levels of \citetalias{pizzolato}.  Clusters of similar age and behaviour are grouped for ease of interpretation.}
\label{BVplot}
\end{figure}

\subsubsection{General trends}
\label{generaltrends}
In Fig.~\ref{BVplot} we plot $\log (L_X/L_{bol})$, where $L_X$ and $L_{bol}$ are the X-ray and bolometric luminosities, against $(B-V)_0$ colour for all of the clusters used in this work along with the saturation levels of \citetalias{pizzolato}.  The 5 young clusters, NGC6530, NGC2547, IC2602, IC2391 and Alpha Persei, show very similar behaviour in bins that contain enough stars to determine a mean $\log (L_X/L_{bol})$ (i.e. one based on more than 3 stars).  As such we group these clusters together as the `young clusters' in Fig.~\ref{BVplot} for clarity.  Given the very young age of NGC6530 and the relatively large age spread of these clusters we consider the young cluster line to be a good first estimate of the X-ray saturation levels across the $(B-V)_0$ range considered in this work.  In the $(B-V)_0$ range covered by \citetalias{pizzolato} this initial approximation to the saturation levels is similar to, if slightly lower than, their levels and continues the trend for the saturation level to decrease for earlier spectral types.  Fig.~\ref{BVplot} shows in broad terms a general trend in the falling off of X-ray emission as a function of time in combination with lower relative X-ray emission of earlier type stars within the same age group.

We note that the Pleiades and NGC2516 have large errors in $\log (L_X/L_{bol})$ in the central bins.  This is the result of a bimodal behaviour of the stars in these clusters within the $(B-V)_0$ range of these bins.  We believe this behaviour is a feature of the age of these clusters and is discussed further in Section \ref{bimodal}.

\begin{figure*}
\includegraphics[width=80mm]{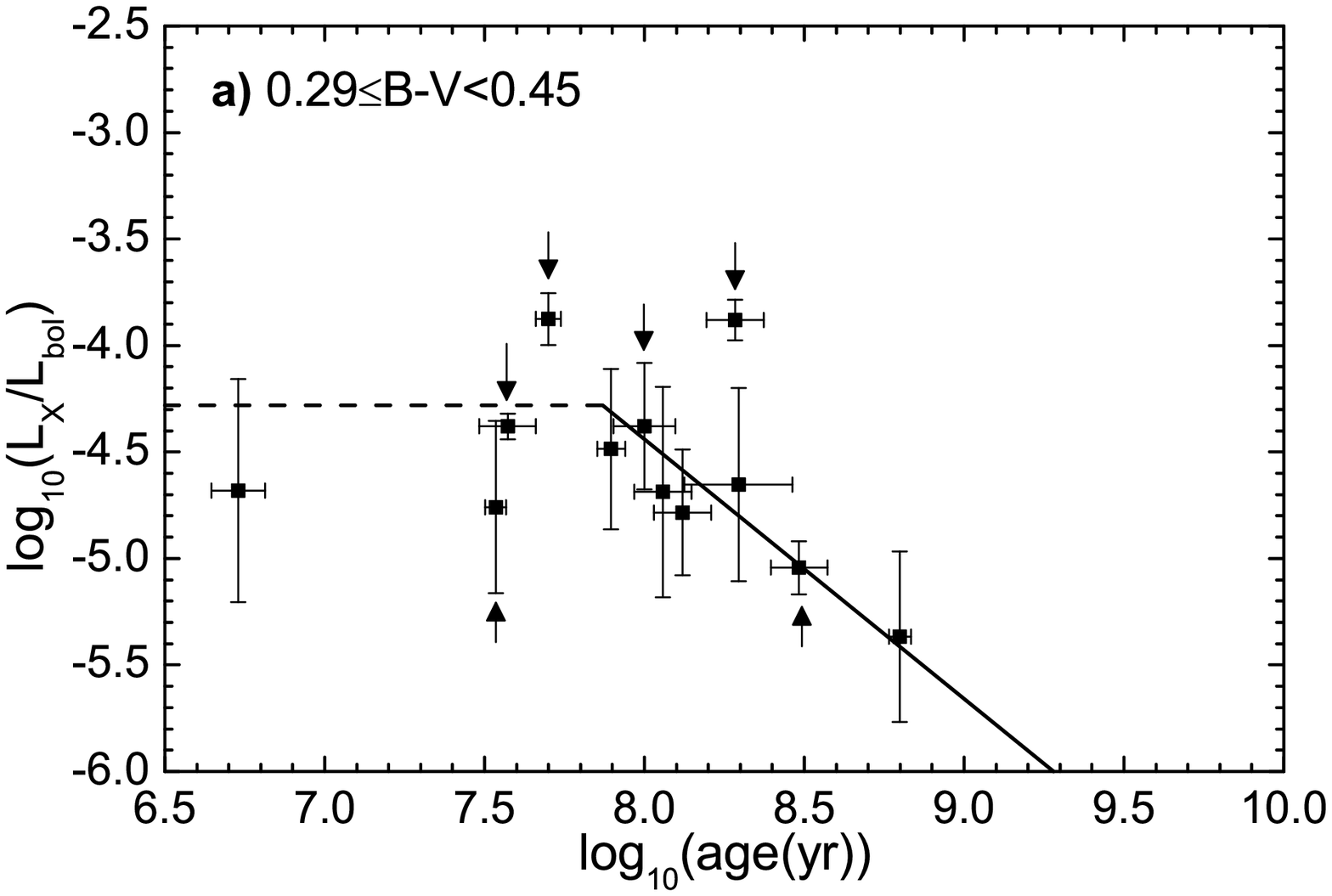}
\includegraphics[width=80mm]{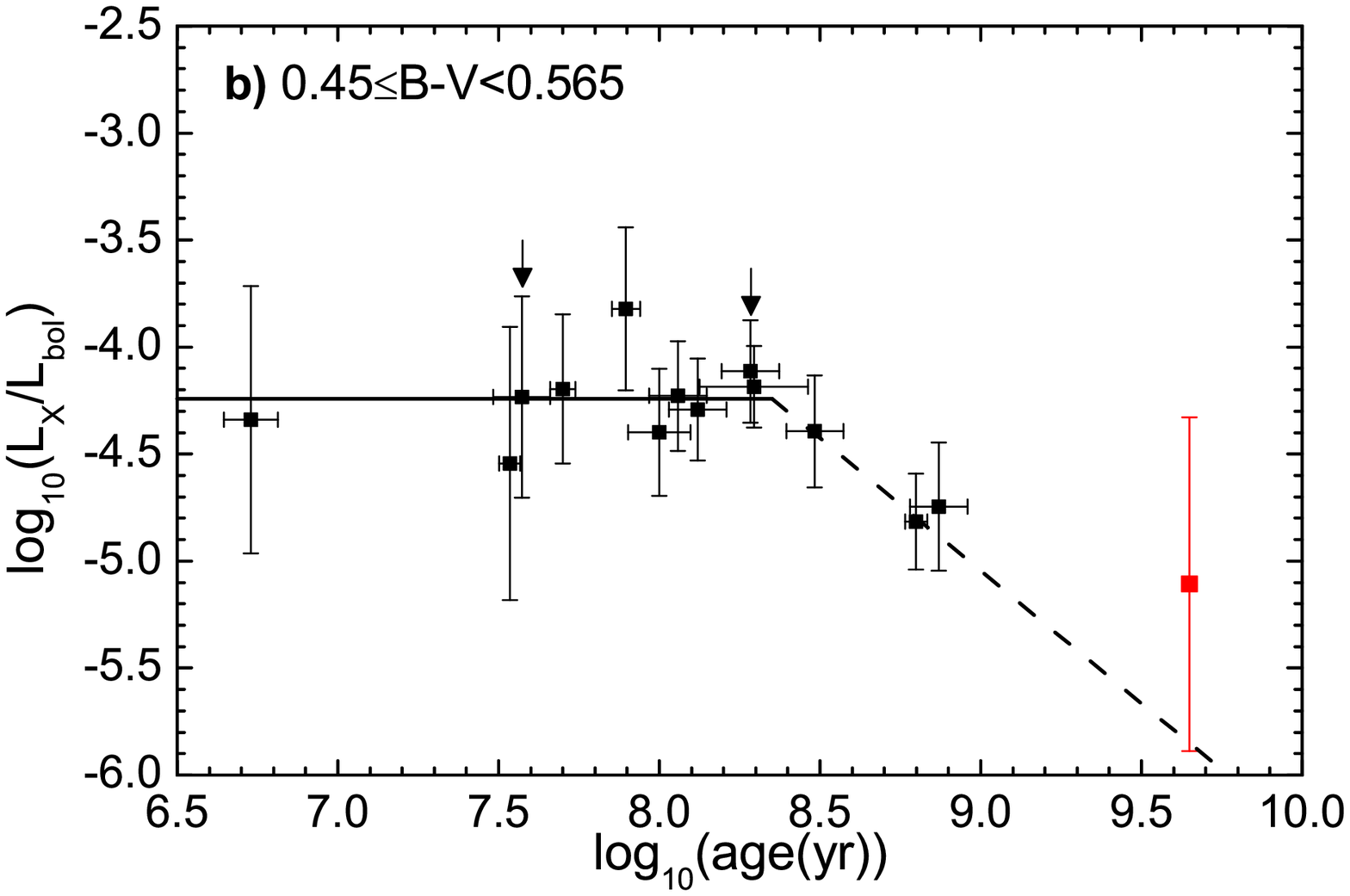}
\includegraphics[width=80mm]{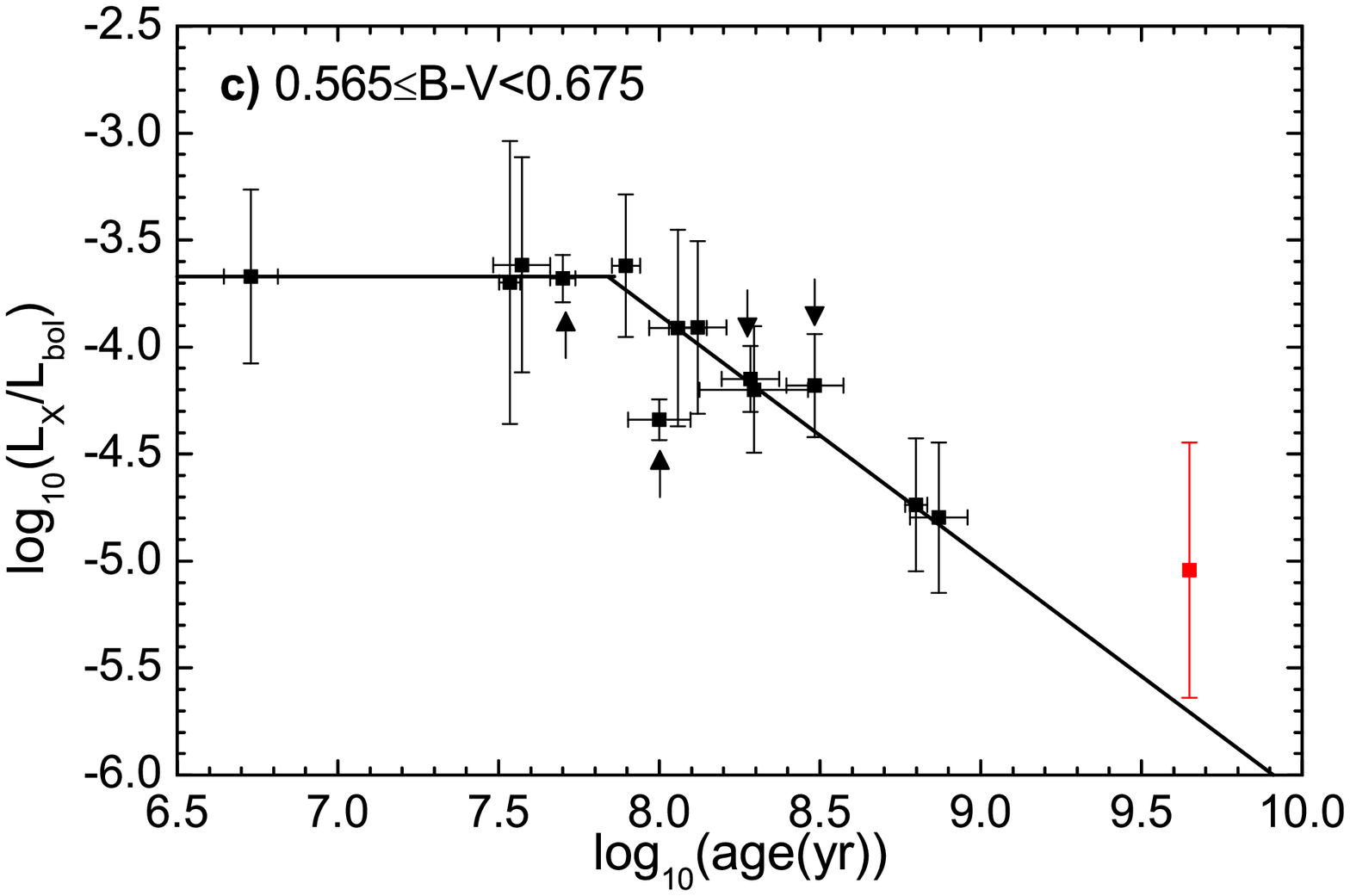}
\includegraphics[width=80mm]{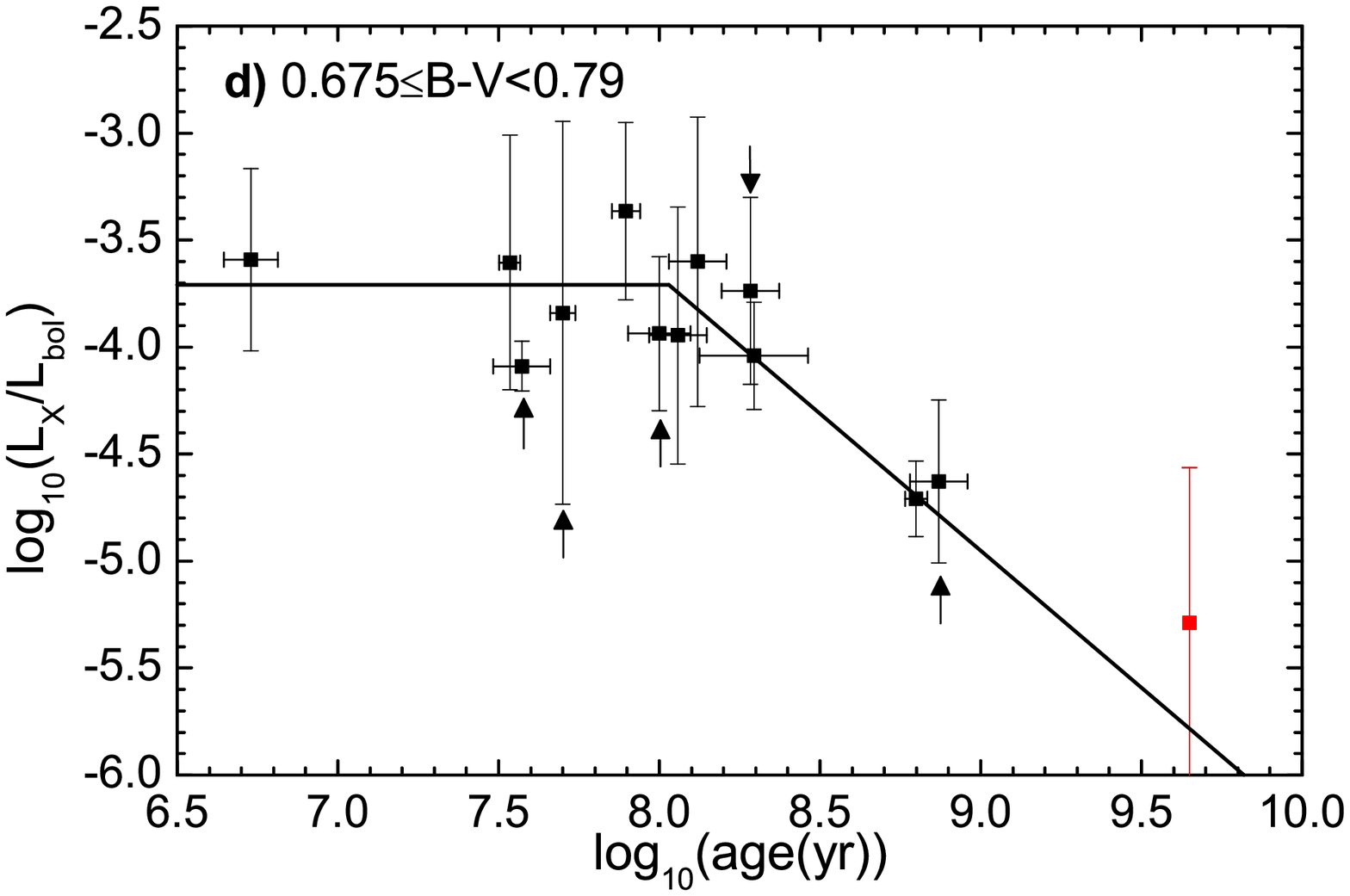}
\includegraphics[width=80mm]{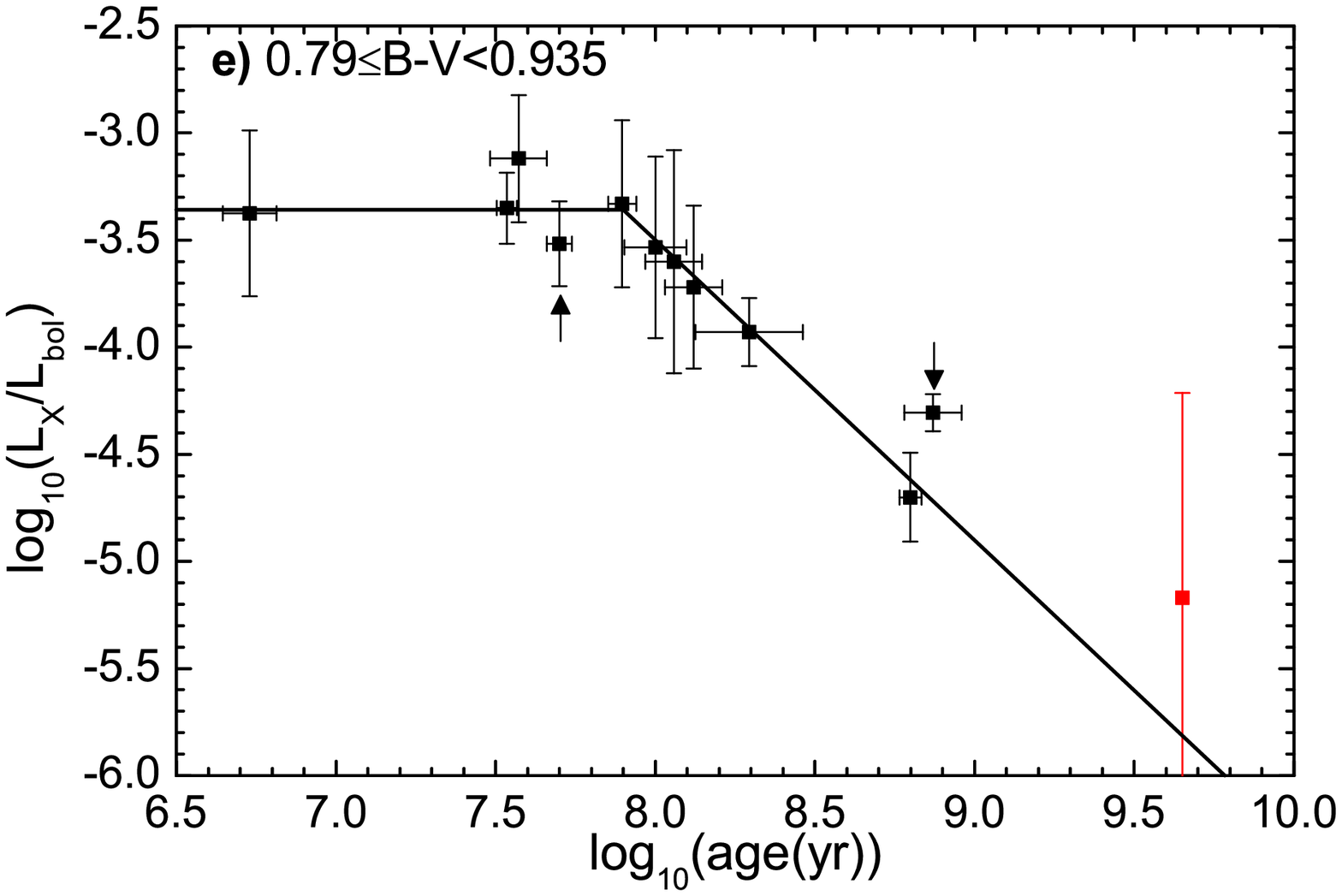}
\includegraphics[width=80mm]{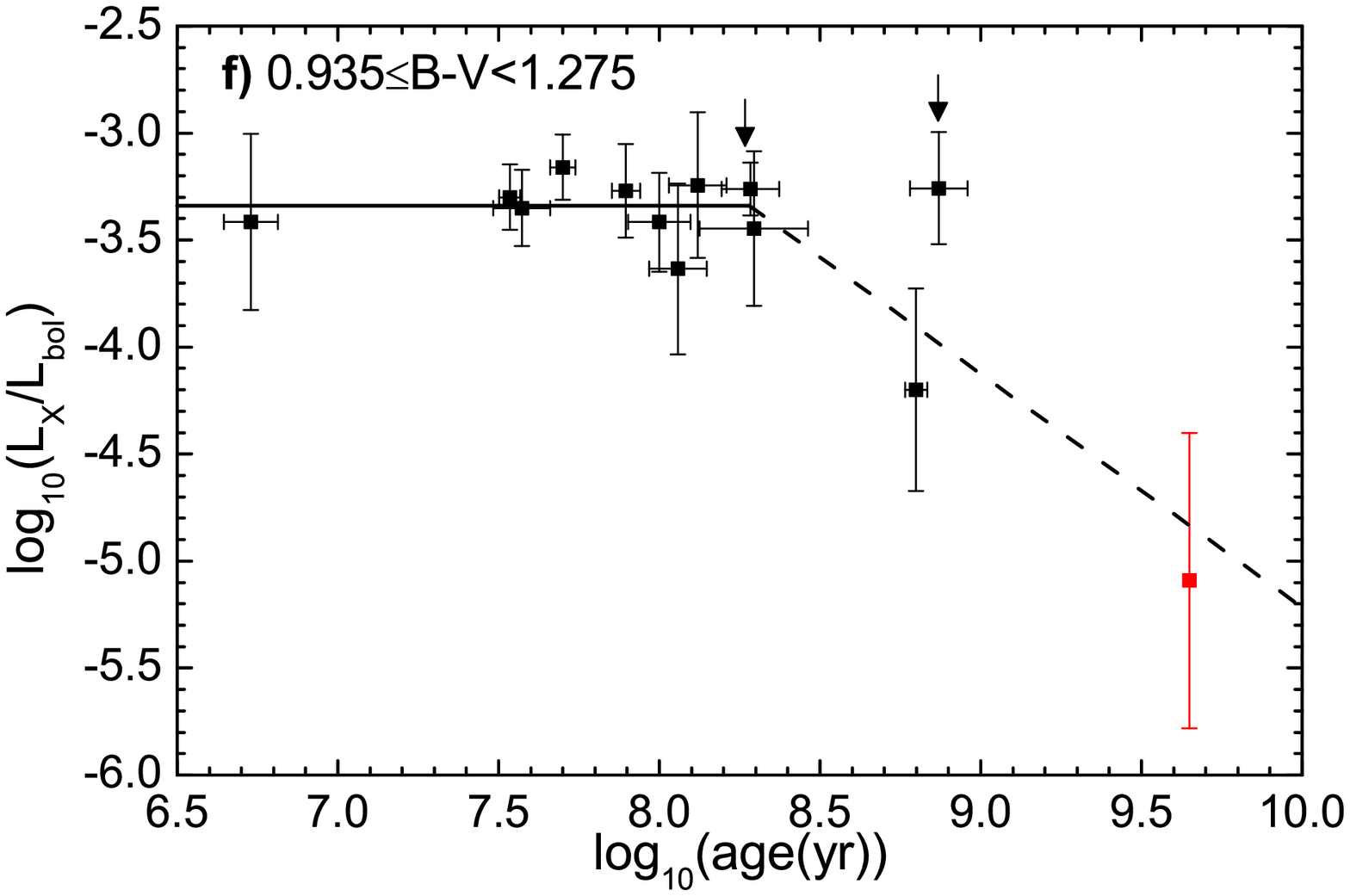}
\includegraphics[width=80mm]{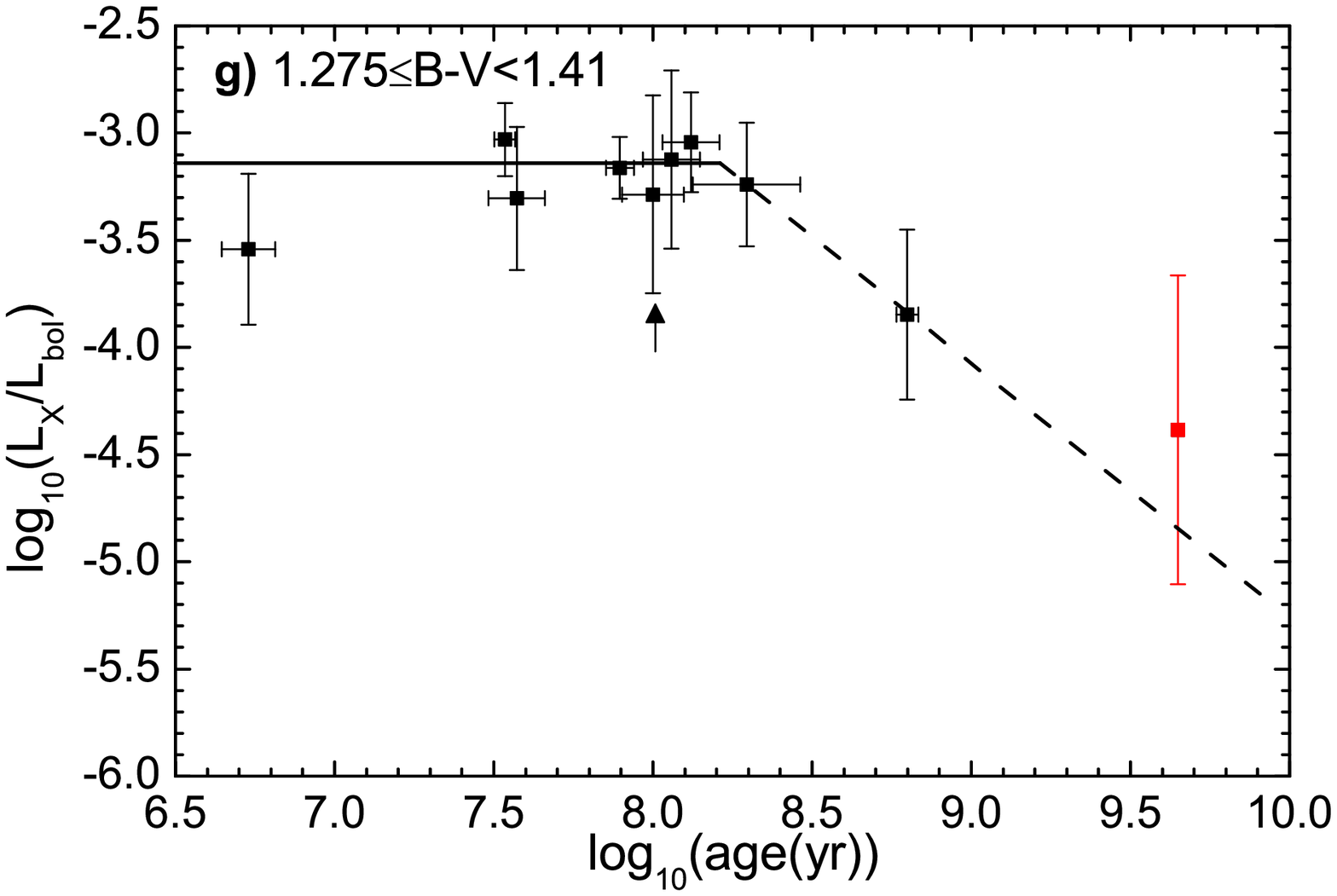}
\caption{X-ray to bolometric luminosity ratio against age for the open clusters used in this work.  Arrows indicate clusters for which 3 or fewer of the selected stars fall within the relevant $(B-V)_0$ bin, these clusters receive a lower weighting in the fitting procedure.  Solid lines indicate the fits to the data with dashed lines as fits that are less certain.  We include the field star sample of \citetalias{pizzolato} (marked in red) at an assumed age of 4.5 Gyr as a guide but do not include them in any of the fitting.  See Sections \ref{regimefits} and \ref{B-Vdependence} for further information.}
\label{XrayBV}
\end{figure*}

\begin{table*}
\begin{minipage}{140mm}
\caption{Results of the fits to the saturated and non-saturated regimes for each $(B-V)_0$ colour bin.  Note that although the $0.935 \leq (B-V)_0 < 1.275$ bin has many more stars than most bins almost half of these are from the same cluster, NGC6530.}
\label{satvalues}
\begin{tabular}{l@{$\leq (B-V)_0 <$}l rcll}
\hline
\multicolumn{2}{c}{$(B-V)_0$ colour range} &	No. of stars &	$\log(L_X/L_{bol})_{sat}$ &	$\log \tau_{sat}$ (yrs) &	Power law index, $\alpha$\\
\hline
0.290&0.450 &	67 &	$-4.28 \pm 0.05 \pm 0.50$ &	$7.87 \pm 0.10 \pm 0.35$ &			$1.22 \pm 0.30$\\
0.450&0.565 &	97 &	$-4.24 \pm 0.02 \pm 0.39$ &	$8.35 \pm 0.05 \pm 0.28$ \hspace{4pt}$(8.30)$ &	$1.24 \pm 0.19$\\
0.565&0.675 &	92 &	$-3.67 \pm 0.01 \pm 0.34$ &	$7.84 \pm 0.06 \pm 0.25$ &			$1.13 \pm 0.13$\\
0.675&0.790 &	79 &	$-3.71 \pm 0.05 \pm 0.47$ &	$8.03 \pm 0.06 \pm 0.31$ &			$1.28 \pm 0.17$\\
0.790&0.935 &	109 &	$-3.36 \pm 0.02 \pm 0.36$ &	$7.90 \pm 0.05 \pm 0.22$ &			$1.40 \pm 0.11$\\
0.935&1.275 &	220 &	$-3.35 \pm 0.01 \pm 0.37$ &	$8.28 \pm 0.07 \pm 0.29$ \hspace{4pt}$(8.27)$&	$1.09 \pm 0.28$\\
1.275&1.410 &	55 &	$-3.14 \pm 0.02 \pm 0.35$ &	$8.21 \pm 0.04 \pm 0.31$ \hspace{4pt}$(8.21)$ &	$1.18 \pm 0.31$\\
\hline
\end{tabular}\\
{Values in parentheses indicate alternative turn-off ages obtained using a fixed gradient of -1 for the fit to the unsaturated regime.}
\end{minipage}
\end{table*}

\subsubsection{The saturated and non-saturated regimes}
\label{regimefits}
\begin{figure}
\includegraphics[width=85mm]{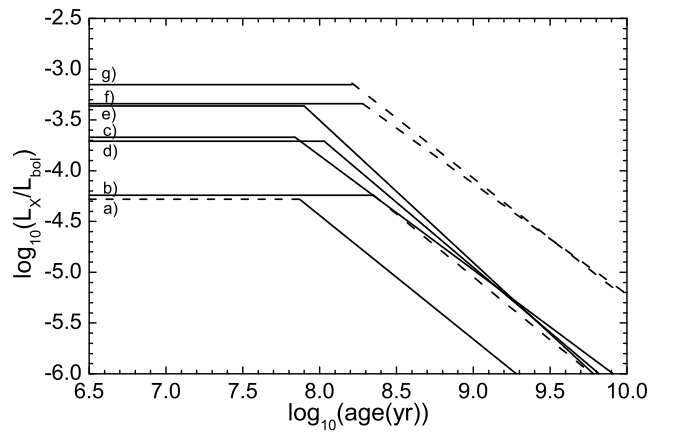}
\caption{Summary of the fits from Fig.~\ref{XrayBV}.  Each line is marked with the letter of the Fig.~\ref{XrayBV} plot from which it is taken.  See Sections \ref{regimefits} and \ref{B-Vdependence} for further information.}
\label{XrayBVsum}
\end{figure}

In Fig.~\ref{XrayBV} we plot the ratio of X-ray to bolometric luminosity against age for all of the clusters in our sample for each $(B-V)_0$ bin along with our fits to the saturated and non-saturated regimes.  The results are summarised in Table \ref{satvalues} and Fig.~\ref{XrayBVsum}.

In each $(B-V)_0$ bin we fit a broken power law to the clusters.  This is done using an iterative implementation of the method described by \citet{fasano1988} with a slight modification to their weighting factor, $W_i$, by the addition of a factor $C_i$ to the numerator.  In each bin the factor $C_i$ is equal to the number of stars in the $i$th cluster in that bin, normalised to the total number of stars in the bin.  This is to account for the fact that clusters with fewer stars in a given bin tend to have a smaller luminosity ratio spread as a result of under-sampling of the distribution.

For the fitting the saturated regime is constrained to be horizontal, while the slope of the unsaturated regime is allowed to vary.  In the case of the $0.45 \leq (B-V)_0 < 0.565$, $0.935 \leq (B-V)_0 < 1.275$ and $1.275 \leq (B-V)_0 < 1.41$ bins however there are relatively few clusters lying in the unsaturated regime.  As such for these bins we also used a fit with the slope of the unsaturated regime fixed at -1.  An unsaturated slope of -1 is what one would expect as a result of the inverse square root decay of rotational frequency with time found by \citet{skumanich} and the $L_X/L_{bol} \propto \omega^2$ relation (where $\omega$ is rotational frequency) found by e.g.~\citetalias{pizzolato} and \citet{pallavicini1981}.  The turn-off ages obtained from these fixed slope fits are included in Table \ref{satvalues} as the values in brackets and are very similar to the turn-off ages obtained from the variable slope fits.

For both the saturation level and saturation turn-off age we quote two errors.  The first error is the error in the position of the saturated level and saturation turn-off age given by the fitting procedure. The second is the root mean square scatter of the individual stars about these mean positions as a measure of the intrinsic scatter in the relations.  The error in the gradient of the unsaturated regime is taken from the fitting procedure similarly to the first saturation level and turn-off age errors and we note that the stellar scatter about the unsaturated regime line is of similar magnitude to the scatter about the mean saturated level.

\begin{figure}
\includegraphics[width=85mm]{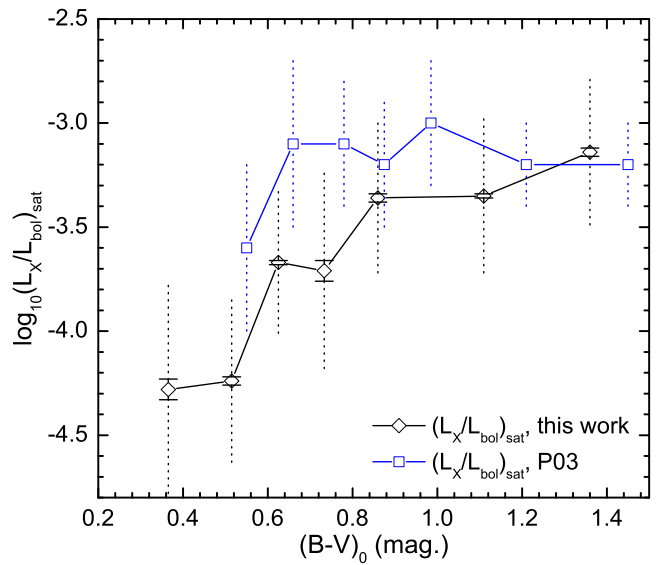}
\caption{Saturated X-ray to bolometric luminosity ratio for our work and that of \citetalias{pizzolato} against $(B-V)_0$ colour.  The capped error bars for our saturated luminosity ratio correspond to the error in the placement of the mean saturated level with the dotted capless error bars corresponding to the typical spreads about the saturated level.}
\label{ratioBV}
\end{figure}

\begin{figure}
\includegraphics[width=85mm]{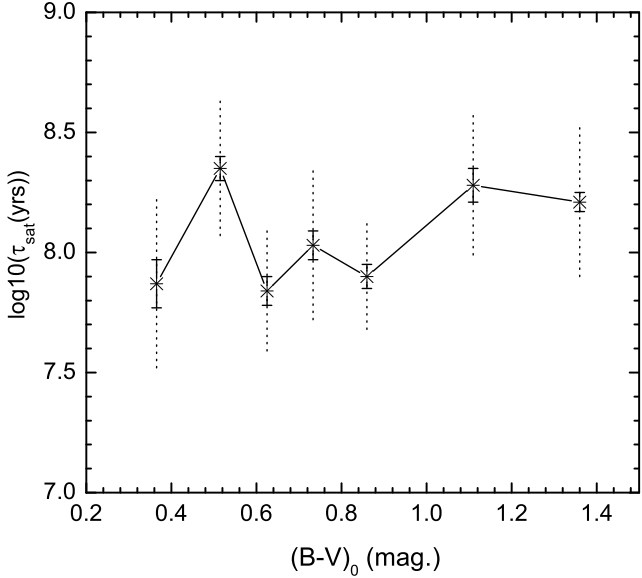}
\caption{Saturation turn-off age against $(B-V)_0$ colour.  The capped error bars correspond to the error in the fitting of the position of the saturation turn-off age with the capless error bars indicating the typical spread about the turn-off age.}
\label{timeBV}
\end{figure}

\subsubsection{$(B-V)_0$ dependence of the saturated and non-saturated X-ray emission}
\label{B-Vdependence}
In Fig.~\ref{ratioBV} we plot both our saturated luminosity ratios and those of \citetalias{pizzolato} for comparison.  Our saturated luminosity ratios are in reasonable agreement with those found by \citetalias{pizzolato}, if in general slightly lower, and fall off consistently as one moves to lower $(B-V)_0$.  We would expect our saturation levels to be slightly lower than those of \citetalias{pizzolato} since they specifically select stars that are saturated and are constrained by requiring rotation periods for the stars.  Since stars will be born with some finite spread of rotation rates however there will be (particularly near the saturation turn-off age) some stars included in the saturated regime that are rotating more slowly than their cluster companions.  These stars will become unsaturated sooner than the rest of the stars in the cluster, the effect of which would be to lower the apparent saturation level slightly.  Additionally rotation periods are generally easier to obtain for the most active stars.

Fig.~\ref{timeBV} shows the saturation turn-off ages obtained against $(B-V)_0$.  Here, unlike for the saturated luminosity ratio there is no obvious trend of the saturation turn-off age with $(B-V)_0$.  Taking the whole FGK range the behaviour of the saturation turn-off age is consistent with a scatter around an age of $\sim$100 Myr.

The gradient of the linear fit to the non-saturated regime gives us $\alpha$, the index in the power law $(L_X/L_{bol})=(L_X/L_{bol})_{sat}(t/\tau_{sat})^{-\alpha}$, where $t$ is the age of the star and $\tau_{sat}$ is the saturation turn-off time for that class of star.  $-\alpha$ is thus simply the slope of the unsaturated regime.  We find that the mean value of $\alpha$ is $1.22 \pm 0.10$ and there is no obvious trend in the values of $\alpha$ either across the whole $(B-V)_0$ range of this work or ignoring the earliest 2 bins to account for any possible dynamo transition.  This is slightly larger than the $\alpha=1$ that we would expect for a $t^{-0.5}$ evolution of rotation frequency with time coupled with an $L_X/L_{bol} \propto \omega^2$ dependence first suggested by \citet{skumanich}.  However we note that more recent studies, e.g.~\citet*{soderblom1}, \citet{pace}, \citet{preibisch} find faster rates for the rotational frequency evolution, mostly in the range $t^{-0.65}-t^{-0.75}$, which is similar to the $t^{-0.61}$ relation we find assuming an $L_X/L_{bol} \propto \omega^2$ dependence.  Additionally there is a degree of degeneracy between $\alpha$ and the saturation turn-off age as the adjustment to the fits of Fig.~\ref{XrayBV} that produces the minimal change to the goodness-of-fit is one which increases or decreases both the saturation turn-off age and $\alpha$ simultaneously.

Ideally we would include X-ray data from clusters older than the Hyades to better constrain the behaviour of the unsaturated regime.  X-ray studies of such clusters are few however and suffer from low detection rates due to the lower fractional X-ray luminosities of older stars.  As mentioned in Section \ref{starsample} we include the sample of field stars used by \citetalias{pizzolato} assuming an age comparable with solar at $\sim4.5$ Gyrs on our plots in Fig.~\ref{XrayBV} (though this sample is not used in any of the fitting) to give a rough guide as to how reasonable it is to extend our result for the unsaturated regime to older stars.  For the most part our unsaturated regime fit passes below the mean of the field stars, though always within the scatter in the sample.  As such we consider that, while caution should be taken, our unsaturated regime results can reasonably be extended to stars of solar age.  The rather large scatter in the field star sample is in part likely a result of the assumption of a uniform age of 4.5Gyrs for the sample.  Within a cluster all of the stars will have very similar ages, albeit that this age will have an uncertainty. On the other hand for the field stars we would expect there to be a real, and probably significant, spread in age and thus that the field stars would in fact be expected to occupy an elliptical region aligned with the unsaturated regime.  Unfortunately the ages of field stars is very difficult to estimate.  We do however note the study of \citet*{garces2011} who use wide binaries consisting of a white dwarf and a K-M type star combined with white dwarf cooling models to produce a sample of old stars with well defined ages.  Also note that since we include earlier type stars in our study than that of \citetalias{pizzolato} there are no field stars in the lowest bin.

Overall our results for the evolution of the X-ray luminosity of FGK stars can be summarised by the equations:
\begin{equation}
L_X/L_{bol} = 
\begin{cases}
	(L_X/L_{bol})_{sat} 				& \text{for } t \leq \tau_{sat},\\
	(L_X/L_{bol})_{sat}(t/\tau_{sat})^{-\alpha} 	& \text{for } t > \tau_{sat}.
\end{cases}
\label{Lxeq}
\end{equation}
where $(L_X/L_{bol})_{sat}$, $\tau_{sat}$ and $\alpha$ are taken from Table \ref{satvalues} for the $(B-V)_0$ range appropriate to the star in question.  Fig. \ref{XrayBVsum} illustrates these relations for each of the $(B-V)_0$ bins we consider in this work.

\subsubsection{Effects of binarity and variability}
\label{binandvar}
As noted in Section \ref{exclusioncriteria} our datasets include several types of variable star, particularly amongst the PMS stars.  The major types of variable star we find in our datasets, that are not excluded, are T Tauri, BY Draconis and flare stars.

T Tauri stars are a class of PMS star that exhibit rapid and unpredictable variability and are likely to be associated with flaring (e.g.~\citealt{appenzeller1989}).  As PMS stars they will also have larger radii than their final MS state and thus be more luminous, and as they lack a hydrogen burning core they will have deep convective zones.  We do not exclude T Tauri stars as their higher luminosity will be taken into account in the calculations and flaring events during observations should be identifiable (see flare stars, below).

BY Draconis variables are late type variables characterised by periodic, rotational, brightness variations due to extensive star spot coverage and longer term modulations due to changing star spot configurations (e.g.~\citealt{alekseev}).  This variability is linked to coronal activity of the same type that leads to X-ray emission and as such we expect a BY Draconis designation to merely be an indicator of, likely saturated, coronal X-ray activity.  Any variability in X-rays due to star spots rotating into and out of view is likely to be fairly low level and so should not have significant effects on the measured X-ray luminosity, though this may contribute to some of the observed scatter.

Flare stars, also known as UV Ceti stars, exhibit short lived, but very powerful outbursts (e.g.~\citealt{tovmassian2003}).  These flares are X-ray bright and can dramatically increase the X-ray luminosity of the star.  Outside a flaring event however the stars are likely to have normal saturated X-ray emission (since flaring is associated with strong coronal activity).  As a large fraction of late type stars in young open clusters (e.g.\ $\sim 50\%$ of Pleiades K type stars) are classified as flare stars we feel that the benefit of the greater number statistics is worth the risk of including an unidentified flaring event.  From \citet{jeffries2} we expect that only a small proportion of stars that are identified as regularly undergoing flares will be in flare during any one observation.  Where data exists on flares that occured during the observations we utilised this to remove stars that were definitely in flare.  In addition (as indicated in Section \ref{exclusioncriteria}) some proportion of flaring events can be identified purely by the extreme X-ray brightening of the star independent of detailed flare studies.

A substantial proportion of the stars in our sample are likely to be binaries or multiples and clearly this will have some effect on the X-ray to bolometric luminosity ratios obtained.  The $(B-V)_0$ colour of a binary or multiple system will be closest to that of the most massive component, since this will be the brightest optically.  In the case of a binary or multiple with components of similar mass the ratio of the total X-ray to total bolometric luminosity of the system should be similar to that of the individual stars and so have minimal effect on the distributions.  For systems with components of rather unequal mass, e.g. a pair with types F and K, the hotter star will dominate the optical/bolometric luminosity but the X-ray to bolometric luminosity ratio of the less massive star could be an order of magnitude higher.  The difference in bolometric luminosities however is such that even in an extreme case the X-ray luminosity of the less massive star is unlikely to be greater than that of the more massive star, and is in general unlikely to be greater than half that of the more massive star.  While our study does not cover M-type stars we note that \citetalias{pizzolato} find the saturated luminosity ratio of early-mid M stars to be very similar to that of late K stars.  So while \citetalias{pizzolato} suggest that M stars may have significantly longer saturated periods their very low bolometric luminosities mean we do not expect putative M star companions to have a significant effect.

We thus expect that the largest typical increase in the $\log (L_X/L_{bol})$ of the star/system is $\sim 0.1-0.2$ dex due to binarity, with an increase of up to $\sim 0.3$ dex in extreme cases.  Therefore binarity will not have too great an effect on the mean $\log (L_X/L_{bol})$, though there may be a small systematic raising.  As we discussed in Section \ref{exclusioncriteria} in contact/interacting binaries the interaction between the stars is likely to have significant effects on the X-ray properties of the stars and while we exclude such systems where they have been identified there may be unidentified interacting binaries in the data.  The number of interacting binaries is likely to be much lower than the number of non-interacting binaries so this should be a small effect.

\subsubsection{The Pleiades and NGC2516}
\label{bimodal}
The Pleiades and the similarly aged NGC2516 exhibit a bimodal population with respect to $\log (L_X/L_{bol})$ in the range $0.55 \leq (B-V)_0 < 0.85$, with an upper branch lying at the saturated level and a lower branch in an unsaturated regime.  Further details will be given in a subsequent paper but note that the main effect here is simply to increase the error in the mean X-ray to bolometric luminosity ratio in the Pleiades and NGC2516 over this range.  It is plausible that this phenomenon could influence the apparent turn-off age in the $(B-V)_0$ range at which this bimodal behaviour is apparent, especially in the $0.675 \leq (B-V)_0 < 0.79$ bin as this lies at the centre of the bimodal region, but we note that excluding the Pleiades and NGC2516 from the fitting in this bin makes little difference to the turn-off age or unsaturated regime power law obtained.

\section{Exoplanet Evaporation}
\label{exoplanets}
As indicated in Section \ref{introduction} when studying the potential evaporation of close-orbiting exoplanets it is important to understand, and account for, changes in X-ray/EUV irradiation over time.  As such we now apply the results of our study of the evolution X-ray emission from late-type stars to the investigation of the evaporation of exoplanets.

\begin{table*}
\begin{minipage}{175mm}
\caption{Planet data.  Most basic planetary parameters are taken from the Exoplanet Encyclopedia$^a$ with any additional references noted in column (15).  In columns (2)--(7) we list the measured planetary properties, mass ($M_P$), radius ($R_P$), orbital period, semi-major axis ($a$), eccentricity and absolute spin-orbit misalignment angle ($|\lambda|$).  The surface gravity, mean density ($\rho$), mean orbital distance $(\langle a \rangle)$, Roche lobe mass loss enhancement factor $(1/K(\epsilon))$ and planetary binding energy ($PE$) (columns (8)--(12)) are calculated directly from the basic planetary and/or host parameters.  See Sections \ref{evapmodel} for the form of $\langle a \rangle$ and $K(\epsilon)$ and \ref{PEroche} for $PE$, note that the value for $PE$ listed here does not include the Roche lobe correction.  Full Table available online.}
\label{planetext}
\begin{center}
\begin{tabular}{l r@{.}l r@{.}l r@{.}l l l r r rc c c c}
\hline
 &\multicolumn{2}{c}{}&\multicolumn{2}{c}{}&\multicolumn{2}{c}{}& & & &surface & & & & & \\
Planet &\multicolumn{2}{c}{$M_P$}&\multicolumn{2}{c}{$R_P$}&\multicolumn{2}{c}{period}& \hspace{5pt}$a$ & ecc. & $|\lambda|$\hspace{4pt} &gravity& $\rho_P$\hspace{10pt} & $\langle a \rangle$ & $1/K(\epsilon)$ & $\log (-PE)$ & ref. \\
 &\multicolumn{2}{c}{($M_J$)}&\multicolumn{2}{c}{($R_J$)}&\multicolumn{2}{c}{(days)}& (AU) & & (deg.) &(m/s)\hspace{2pt}&(kg/m$^3$)& (AU) & & (J) & \\
(1) &\multicolumn{2}{c}{(2)}&\multicolumn{2}{c}{(3)}&\multicolumn{2}{c}{(4)}& \hspace{3pt}(5) & (6) & (7)\hspace{4pt} &(8)\hspace{4pt} & (9)\hspace{4pt} & (10) & (11) & (12) & (13)\\
\hline
CoRoT-1 b & 1 & 030 & 1 & 490 & 1 & 5090 & 0.0254 & 0 & 77 & 12.02 & 413 & 0.0254 & 2.27 & 36.26 & f\\
CoRoT-10 b & 2 & 750 & 0 & 970 & 13 & 2406 & 0.1055 & 0.53 & -- & 75.73 & 3997 & 0.1203 & 1.06 & 37.30 & \\
CoRoT-11 b & 2 & 330 & 1 & 430 & 2 & 9943 & 0.0436 & 0 & -- & 29.52 & 1057 & 0.0436 & 1.38 & 36.99 & \\
CoRoT-12 b & 0 & 917 & 1 & 440 & 2 & 8280 & 0.0402 & 0.07 & -- & 11.46 & 407 & 0.0403 & 1.62 & 36.18 & \\
CoRoT-13 b & 1 & 308 & 0 & 885 & 4 & 0352 & 0.0510 & 0 & -- & 43.27 & 2503 & 0.0510 & 1.20 & 36.70 & \\
\hline
\end{tabular}\\
\end{center}
\emph{References}: a)~www.exoplanet.eu, b)~\citet{buchhave2011}, c)~\citet{gillon}, d)~\citet{johnson2011}, e)~\citet{narita2010}, f)~\citet{pont2010}, g)~\citet{queloz2010}, h)~\citet{triaud2009}, i)~\citet{winn2010a}, j)~\citet{winn2010b}, k)~\citet{winn2011}\\*
\end{minipage}
\end{table*}

\subsection{Exoplanet sample}
\label{planetsample}
We select transiting planets for our study as we require the planetary radius in order to calculate the surface gravity and investigate the evaporation rates.  We selected our planet sample from the Exoplanets Encyclopaedia\footnote{www.exoplanet.eu, see also \citet{schneider2011}} on 13 September 2011 and take the majority of the planetary parameters required for our analysis from the encyclopaedia.  Where other sources were consulted these are listed in the online material along with the planet data.

While we seek to use as large a sample as possible there are a number of systems that lack some of the data required for our analysis.  As such from the initial sample of 176 transiting exoplanets we remove the KOI-703 system, the latest tranche of WASP systems (20, 42, 47, 49 and 52 through 70) and WASP-36b as many of the characteristics of the hosts and systems are not yet well defined.  We remove GJ1214b and GJ436b as these have M type host stars and WASP-33b as this has an A type host, and thus fall outside the scope of this study.  HAT-P-31b, Kepler-5b, 6b, 15b, OGLE-TR-182b, SWEEPS-04b, 11b, TrES-5b, WASP-35b and 48b are removed as the spectral type of the host star is not well defined.  Kepler-10c, 11g and 19b are removed as the mass of the planet is not well constained and we remove CoRoT-21b, HD 149026b, Kepler-7b, 8b, KOI-423b and KOI-428b as they have evolved hosts.

In addition to systems where some of the necessary data is lacking the improvements in transit surveys are now making the regime of super-Earth/sub-Neptune class planets accessible.  This means there are some low-mass, rocky, planets for which our evaporation model, based on a hydrogen rich composition, would not be appropriate.  CoRoT-7b, 55 Cnc e, HD97658b, Kepler-10b and planets b, c, d, e and f in the Kepler-11 system are all less massive than Uranus and Neptune.  However models of the interiors of super-Earths (e.g. \citealt{seager2007, sotin2007, wagner2011}) show that, with the exception of Kepler-10b and CoRoT-7b, all of these planets have bulk densities too low for a primarily silicate rock composition.  These studies suggest that for Kepler-11b and 55 Cnc e a water-rich composition similar to that of Ganymede with $\ga 50\%$ water by mass is possible.  HD97658b and Kepler-11c, d, e and f however must have significant gaseous envelopes and these are better thought of as sub-Neptunes than super-Earths.  As such we do not include 55 Cnc e, CoRoT-7b, Kepler-10b or Kepler-11b in the main sample as our evaporation model is unlikely to describe these planets properly.  We will however discuss 55 Cnc e, CoRoT-7b and Kepler-10b further later as they have the potential to be the evaporated cores of gas giants (e.g.~\citealt{bjackson2010, valencia2010}), termed `chthonian' planets by \citet{lecavelier2004}, and so may have been influenced by the same evaporation processes in the past.  These three super-Earth class planets are listed separately at the bottom of Tables~\ref{planetext} and \ref{hostext}.

Of the remaining planet hosts HAT-P-14, 15, OGLE-TR-10, 111, 113 and TrES-3 and 4 lacked spectral type determinations but had $B-V$ data available; in these cases we estimated spectral types from the $B-V$ values.  We expect reddening to be negligible for the HAT and TrES planets as these are relatively nearby.  This is not the case for the OGLE planets though so for these we may have estimated a later spectral type than the reality.  As our spectral type/$(B-V)_0$ bins for the X-ray investigation are relatively wide and the bolometric correction is not rapidly varying with $B-V$ we do not expect errors in the spectral type estimation to have large effects however.  This then leaves us with a sample of 121 planets, which represents a substantial extension to the sample used in \citetalias{davis}, a result of the accelerating rate of exoplanet discoveries.  A full list of the planets used and their parameters, and those of their host stars, can be found in the online material\footnote{online material web address} with extracts given in Tables \ref{planetext} and \ref{hostext} as a guide.

\begin{figure}
\includegraphics[width=80mm]{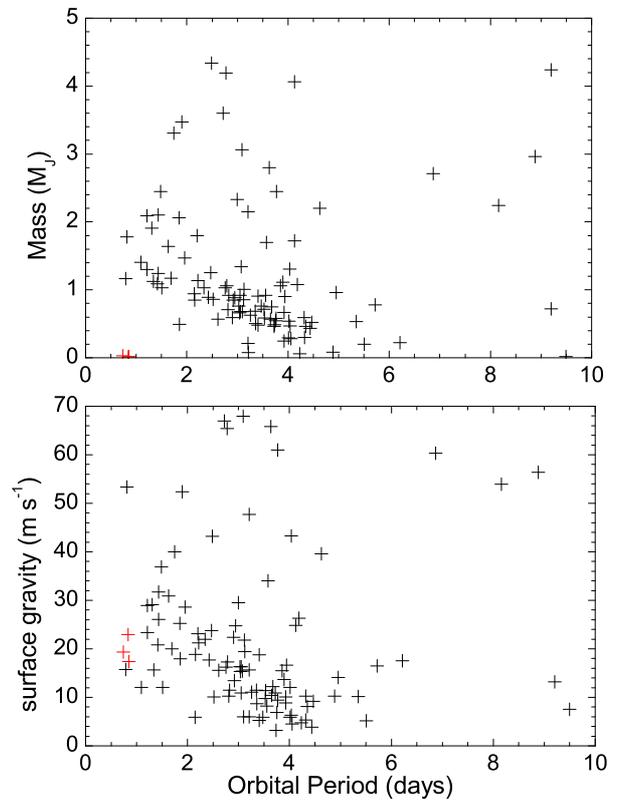}
\caption{Plots of the correlations reported by \citet{mazeh} and \citet{southworth} for the sample of planets used in this work.  Planets exist to the top and right of the plot areas but these regions are sparsely populated.  The positions of the three super-Earths, 55 Cnc e, CoRoT-7b and Kepler-10b, not included in the main sample, are shown in red}
\label{sgmassperiod}
\end{figure}

In Fig.~\ref{sgmassperiod} we plot mass and surface gravity against orbital period for our exoplanet sample after e.g. \citet{mazeh}, \citet{southworth} and \citetalias{davis}.  There is greater scatter in both of our plots than in the original suggested correlations, particularly to the top and right.  The lower left of both distributions still displays a notably sparse population however despite strong selection biases towards detecting planets that would lie in these regions.  The mass-period distribution in particular displays quite a sharp cut-off at the lower left.  The upper right regions of both plots are also rather sparsely populated, but transit surveys select against long orbital periods and high surface gravities.  Any residual deficit of detections in these regions is unlikely to be related to evaporation but may be a result of planetary migration or formation.

\begin{table*}
\begin{minipage}{175mm}
\caption{Host star parameters.  Most original data are taken from the Exoplanet Encyclopedia$^1$ with any additional references listed in column (9).  Where a $B-V$ colour is listed in column (3) the corresponding spectral type has been determined from the colour using the tables presented in \citet{lang}.  The bolometric magnitude is calculated from the $V$-band magnitude and a bolometric correction based on the spectral type (also using the tables of \citealt{lang}).  $E_X$ is the total X-ray energy emitted by the star over a period from formation to its present age as calculated using our X-ray characteristics for the appropriate spectral type as presented in Table \ref{satvalues} and described in Sections \ref{regimefits} and \ref{B-Vdependence}.  Full Table available online.}
\label{hostext}
\begin{center}
\begin{tabular}{llc r r r r @{}l r @{}l l}
\hline
Star & mass & spectral & $V$ mag. & bol. mag. & distance  & Age && $\log (E_X)$ &  & Refs.\\
 & ($M_{\odot}$) & type  & (mag.) & (mag.) & (pc)\hspace{5pt} & (Gyr) && (J)\hspace{8pt} &  & \\
(1) & \hspace{5pt}(2) & (3)  & (4)\hspace{6pt} & (5)\hspace{6pt} & (6)\hspace{6pt} & (7) && (8)\hspace{8pt} &  & (9)\\
\hline
CoRoT-1 & 0.95 & G0V & 13.6 & 13.42 & 460 & 4 & + & 38.73 & & \\ 
CoRoT-10 & 0.89 & K1V & 15.22 & 14.85 & 345 & 3 &  & 38.12 & & \\ 
CoRoT-11 & 1.27 & F6V & 12.94 & 12.8 & 560 & 2 &  & 38.9 & & \\ 
CoRoT-12 & 1.08 & G2V & 15.52 & 15.32 & 1150 & 6.3 &  & 38.79 & & \\ 
CoRoT-13 & 1.09 & G0V & 15.04 & 14.86 & 1310 & 1.6 &  & 39 & & \\
\hline
\end{tabular}\\
\end{center}
\emph{References}: 1) www.exoplanet.eu, 2) \citet{alsubai2011}, 3) \citet{anderson2010}, 4) \citet{anderson2011}, 5) \citet{bakos}, 6) \citet{chan2011}, 7) \citet{colliercameron2007}, 8) \citet{desert2011}, 9) \citet{hartman}, 10) \citet{hartman2011}, 11) \citet{hartman2011b}, 12) \citet{havel2011}, 13) \citet{hebb2010}, 14) \citet{hellier2011}, 15) \citet{howard2010}, 16) \citet{kovacs2007}, 17) \citet{latham}, 18) \citet{mandushev2007}, 19) \citet{maxted2010}, 20) \citet{maxted2010b}, 21) \citet{melo2006}, 22) \citet{noyes2008}, 23) \citet{odonovan2007}, 24) OGLE (ogle.astrouw.edu.pl), 25) \citet{pal}, 26) \citet{shporer2009}, 27) \citet{skillen}, 28) \citet{smalley2011}, 29) \citet{sozzetti2009}, 30) \citet{torres2010}, 31) \citet{udalski2008}\\*
*$V$ mag.\ or distance information is lacking for these stars and thus $E_X$ was calculated using the bolometric luminosity of a typical star of the same spectral type as the host (taken from the tables presented in \citealt{lang}).\\*
$^+$ No literature age is available for this star so an age of 4~Gyrs is assumed.
\end{minipage}
\end{table*}

\subsection{Evaporation Model}
\label{evapmodel}
All exoplanets will experience evaporation as a result of Jeans escape of material in the high velocity tail of the thermal velocity distribution.  It should be noted that the temperature here that is the determinant of the mass-loss rate is not the effective thermal emission temperature of the planet ($T_{eff}$), but the temperature of the exosphere ($T_{\infty}$), as this is the atmospheric layer from which the escape will be taking place.  $T_{\infty}$ is typically much greater than $T_{eff}$ and for the hot Jupiters that we are considering will be determined almost exclusively by the incident X-ray/EUV flux.  \citet{lammer} showed that for exospheric temperatures likely present on hot Jupiters the mean thermal velocities of the exospheric gas becomes comparable to the planetary escape velocity and thus that the Jeans formulation is no longer appropriate.  They suggested instead that energy limited escape due to absorption of X-ray/EUV radiation in the upper atmosphere is the best model for these hydrodynamic blow-off conditions.  \citetalias{lammer2009} constitutes the most recent and comprehensive study based on this evaporation model, however, as in earlier studies, they rely on relatively coarse estimates of stellar X-ray evolution.

The model presented by \citet{lecavelier1} for the estimation of evaporation rates of exoplanets due to X-ray/EUV flux, and subsequently used by \citetalias{davis}, \citetalias{lammer2009} and others, is shown below as Eq.\ref{evapequation}. This model uses the ratio of planetary binding energy and orbit-averaged X-ray/EUV power incident on the planet to predict the mass loss rates as:
\begin{equation}
\dot{m}=\eta\frac{\displaystyle{L_X R^3_P}}{\displaystyle{3 G M_P \langle a \rangle ^2}},
\label{evapequation}
\end{equation}
Where $\dot{m}$ is the mass loss rate, $L_X$ is the X-ray/EUV luminosity of the star, $M_P$ and $R_P$ are the mass and radius of the planet, $G$ is the gravitational constant, $\langle a \rangle$ is the time averaged orbital distance (given by $\langle a \rangle = a (1 + \frac{1}{2} e^2)$ , with $a$ the semi-major axis and $e$ the eccentricity) and $\eta$ is an efficiency factor.  The gravitational binding energy used in this model assumes that gas giant planets have the density profile of an $n=1$ polytrope.

As indicated above this is an energy-limited model and thus will fail in very low X-ray/EUV irradiation regimes where hydrodynamic blow-off conditions no longer apply and the escape rate calculation should revert to the Jeans formulation.  \citet{erkaev} showed that for a Jupiter mass exoplanet orbiting at $\ga0.15$ AU the exospheric temperature may not be high enough to support hydrodynamic blow-off.  Very few of the planets in our sample lie at orbital distances this large and those that do will, as a result of their larger orbital distances, lie at low mass loss rates.  As such we will assume that all of the exoplanets in our sample are in the hydrodynamic blow-off regime and simply note that the (already low) mass loss estimates for planets with orbital distances $\ga0.15$ AU may be overestimates.

\subsubsection{Roche lobe effects}
\label{rochelobe}
This simple model assumes that a particle must be moved to infinity to escape the gravitational potential of a planet, whereas the true criterion is that the particle must be moved to the edge of the Roche lobe (see Section 4.3 of \citealt{lecavelier1}) and as a result this model will be an underestimate of the true mass loss rate.  \citet{erkaev} showed that for the high mass ratio case of a planet orbiting a host star the gravitational potential of the planet is adjusted by a factor that can be approximated as:
\begin{equation}
K(\epsilon)=1-\frac{3}{2\epsilon}+\frac{1}{2\epsilon^3},
\label{Kepsilon}
\end{equation}
thus enhancing the mass loss rate by a factor $1/K(\epsilon)$ due to the closeness of the Roche lobe boundary to the planet surface, where $\epsilon$ is a dimensionless parameter characterising the Roche lobe boundary distance and is given by:
\begin{equation}
\epsilon=\frac{R_{roche}}{R_P}=\langle a \rangle \left( \frac{4\pi\rho_P}{9M_*} \right)^{\frac{1}{3}},
\label{epsilon}
\end{equation}
With $\rho_P$ the planet density and $M_*$ the mass of the host star.  Including this correction due to Roche lobe effects the mass loss equation then becomes:
\begin{equation}
\dot{m}=\eta\frac{\displaystyle{L_X R^3_P}}{\displaystyle{3 G M_P \langle a \rangle ^2 K(\epsilon)}}.
\label{evapequationroche}
\end{equation}

As we approach $\epsilon$ = 1, $K(\epsilon) \to 0$, and thus the mass loss rate enhancement becomes infinite since the planet is now undergoing dynamical Roche lobe overflow and there is very little dependence on the X-ray/EUV irradiation.  For the present day conditions of the exoplanets in our sample the smallest values of $\epsilon$ are $\sim$1.9, corresponding to a maximum mass loss rate enhancement of $1/K(\epsilon)\sim 3.5$, with the mean value being $1/K(\epsilon)\sim 1.4$.  Thus we expect that Eq. \ref{evapequation} alone will on average underestimate the true mass loss rate by $\sim$40\%, while in the most extreme cases it may underestimate the mass loss rate by a factor of $\sim$3.5.  We will discuss the implications of the mass loss enhancement due to Roche lobe effects further in a later section.

\subsubsection{Evaporation efficiency}
\label{efficiency}
Determination of the efficiency factor $\eta$ is a very complex and, for exoplanets for which the atmospheric composition is not well known, a poorly constrained task.  \citetalias{lammer2009} discuss ranges of possible values of $\eta$ and the contributions to it, settling on a value of $0.1-0.25$.  On the other hand \citet{ehrenreich2011} suggest rather higher values of $\eta$ based on the present mass loss rates of HD209458b and HD189733b.  While it is highly likely that the real value of $\eta$ is somewhat less than 1, given the poor constraints we, as far as possible, do not build a specific value of $\eta$ into our study (thus implicitly using $\eta =1$ as a base from which it is easiest to move to different values of $\eta$).  We do however discuss the effects of different possible values of $\eta$ in our final analysis and provide predictions of mass loss histories for both $\eta=1$ as the maximal, base, case and $\eta=0.25$ as a probable realistic value.

\subsubsection{Evolution of planetary radius}
\label{radiusevol}
A further issue that must be considered is evolution of the planetary radius with time since the planetary radius comes into Eq.~\ref{evapequation} both from the area over which the planet absorbs incoming X-ray/EUV radiation and the gravitational binding energy of the planet.  An unperturbed gas-giant will slowly contract, however for close-orbiting planets there are a number of complications.  The most important in the context of our study is if the planet is undergoing significant mass loss.  \citet{baraffe} study evolution of planetary radii over time under various mass loss regimes and initial planet compositions.  Tidal effects can also be significant for close-orbiting planets and is typically parametrised in terms of an additional internal energy source (e.g. \citealt*{miller2009}).  For a non-evaporating planet this causes contraction to stall at a point determined by the magnitude of the tidal energy source.

The application of detailed modelling of the evolution of planetary structure with mass loss such as that conducted by \citet{baraffe} to all of the planets in our sample is beyond the scope of this study, as is a detailed treatment of tidal heating.  To arrive at a tractable problem it is necessary to make a simplifying assumption about the evolution of the planetary radius.  As such in this study we consider two cases; that of a constant radius (i.e. no evolution, and a constant absorbing area) and that of a constant density with the radius thus scaling in an easily defined and analytically integrable way as the planet loses mass.  Tracing past mass loss back in time from the present state of a planet the constant radius approximation will result in a lower predicted mass loss than the constant density approximation.  This occurs since the younger, more massive, planet will be denser than it is in the present day and so more strongly bound and less easily evaporated.  Conversely when integrating mass loss forward from a fixed starting point the constant radius approximation will predict a higher degree of mass loss.  These two cases in general bracket the results found by \citet{baraffe} and we expect the true mass loss for the majority of the planets in our sample to lie within the two predictions.  The true mass loss will also tend to be closer to the constant radius approximation for planets losing large mass fractions and closer to the constant density approximation for planets losing lower mass fractions.

While we expect the constant radius and constant density approximations to bracket the true mass loss for the majority of planets there may be deviations for planets with the highest and lowest predicted mass loss.  In the late stages of evaporation in which a planet has lost a large fraction of its initial mass \citet{baraffe} find that it can enter a runaway mass loss regime in which the radius expands as it loses mass, accelerating the rate of mass loss.  This runaway regime will not be replicated in our estimates so we note that any planet that loses $>80-90\%$ of its initial mass under our estimates may enter such a runaway regime.  In addition planets with low levels of mass loss would likely undergo some contraction and increase in density as they age, though somewhat less than a true Jupiter analogue due to the effects of tidal heating and thermal (rather than X-ray/EUV) irradiation.  In terms of tracing the mass loss histories of the present population of planets this would result in the degree of mass loss of some planets with low predicted mass loss being slightly underestimated.

\subsection{Integrated X-ray/EUV emission}
\label{integXEUV}
The total X-ray energy emitted over the lifetime of a star, $E^{tot}_X$, will be the sum of the energy emitted during its saturated period and the energy emitted while in the unsaturated regime.  Using the results of Section \ref{Xray} with a saturation turn-off age of $\tau_{sat}$, an unsaturated regime power law index of $-\alpha$ and the present age of the star, $t_0$, the energy emitted while in the saturated and unsaturated regimes, $E^{sat}_X$ and $E^{unsat}_X$ respectively, is:
\begin{equation}
E^{sat}_X = L_{bol} \left( \frac{L_X}{L_{bol}} \right) _{sat} \tau_{sat},
\label{Exsat}
\end{equation}
and
\begin{equation}
E^{unsat}_X = \frac{1}{\alpha-1}L_{bol} \left( \frac{L_X}{L_{bol}} \right) _{sat} \tau_{sat} \left[1- \left( \frac{\tau_{sat}}{t_0} \right)^{\alpha-1} \right],
\label{Exunsat}
\end{equation}
The total X-ray energy emitted over the lifetime of the star, $E^{tot}_X=E^{sat}_X+E^{unsat}_X$, will thus be:
\begin{equation}
E^{tot}_X = \frac{1}{\alpha-1}L_{bol} \left( \frac{L_X}{L_{bol}} \right) _{sat} \tau_{sat} \left[\alpha- \left( \frac{\tau_{sat}}{t_0} \right)^{\alpha-1}\right],
\label{Extot}
\end{equation}
Note that in these equations we implicitly assume that the bolometric luminosity of the star is constant through time and equal to the present bolometric luminosity.  Early in the life of the star before it reaches the main sequence this is evidently not true and PMS stars have significantly higher luminosities than their MS counterparts (see e.g.~\citealt{stahler2005}).  This will increase the X-ray energy emitted during the saturated regime above that predicted by Eq.~\ref{Exsat} above.  Indeed if we take the $0.935 \leq (B-V)_0 < 1.275$ bin (stars of roughly spectral type K3-6) stars from the very young cluster NGC6530 have a mean bolometric luminosity of $\sim 1.5 L_{\sun}$.  This falls to around $0.3-0.4 L_{\sun}$ in NGC2547 and to the expected MS values of $\sim 0.2 L_{\sun}$ by around the age of $\alpha$-Persei and certainly reaching their main sequence luminosity by the Pleiades age.  As such we estimate that by neglecting bolometric luminosity evolution of PMS stars $E^{sat}_X$ as given by Eq.\ref{Exsat} may be an underestimate by a factor of $\sim 2$ for the later type stars in our study with a smaller deviation for the earlier type stars that reach the main sequence sooner.  In addition there is an error of a factor of $\sim$1.5-2 as a result of the spreads in $(L_X/L_{bol})_{sat}$ and $\tau_{sat}$.

As a result of the high level of X-ray emission during the saturated period, and the fall off with time of X-ray emission during the unsaturated period, saturated emission comprises a significant fraction of the total lifetime X-ray emission of the star.  For our X-ray study this fraction is typically $\sim$30 per cent.  As the saturated period is a fairly small fraction of the total lifetime of the system we can expect that the majority of evaporation will occur early in the life of a planet.  The implications of this are discussed further in Sections \ref{ageuncertain} and \ref{migration}.

As mentioned in Section \ref{evapmodel} the Extreme Ultraviolet (EUV) may also play a role in driving the evaporation of the atmosphere of a close orbiting gas giant, however we have only the X-ray luminosities.   The EUV derives primarily from the same coronal processes as stellar X-ray emission and so we expect that it will follow the same temporal variations as the X-ray emission.

There are a number of complications when considering the EUV however.  Firstly it is more difficult to quantify the stellar EUV emission since observations in this band are scarce, and made difficult by absorption in the interstellar medium.  \citet{sanzforcada2011} attempt to circumvent this problem using coronal modelling to extrapolate the EUV flux from the X-ray flux of planet hosting stars that have been observed in the X-ray.  They calibrate this with a small sample of nearby stars that have been observed in the EUV.  Unfortunately the vast majority of planet-hosting stars that have been observed in the X-ray are non-transiting systems reducing the information available about the planets.

There are also issues for the evaporation itself since EUV photons are more weakly penetrating than X-ray photons.  Studies of the photoevaporation of protoplanetary discs (e.g. \citealt*{ercolano2009}; \citealt{owen2010}) have shown this means that in the case of strong X-ray driven outflows the EUV does not penetrate far enough to drive further evaporation and instead just heats the outflow.  Due to these issues we concentrate here solely on evaporation induced by stellar X-ray emission, but note that neglecting EUV emission may result in the underestimation of the total energy available to drive evaporation.

\subsection{Mass loss and destruction limits}
\label{massloss}
With these equations for the X-ray energy emitted by the star over time we can integrate the mass loss (Eq.~\ref{evapequation}) to quantify the expected mass loss of the known planets and to obtain destruction limits.  Under the constant radius approximation integrating the mass loss equation back in time from the present conditions of a planet gives:
\begin{equation}
\label{constRmi}
m^2_i - m^2_t = \frac{2}{3} \frac{R^3}{G \langle a \rangle ^2} \eta E_X,
\end{equation}
while under the constant density approximation we obtain:
\begin{equation}
\label{constdensmi}
m_i - m_t = \frac{1}{4 G \pi \rho \langle a \rangle ^2} \eta E_X,
\end{equation}
where $m_i$ is the initial mass of the planet, $m_t$ is the mass today, $R$ and $\rho$ are the radius and density of the planet (constant under the respective approximations) and $E_X$ is the X-ray energy emitted by the host star over the appropriate time interval, in the case of integrating over the entire lifetime this will be $E^{tot}_X$.

\subsubsection{Destruction limits in the $M^2_P/R^3_P$ vs $\langle a \rangle^{-2}$ plane}
\label{destlimits}
We can obtain `destruction lines' from equations \ref{constRmi} and \ref{constdensmi} by setting $m_t =0$, i.e.\ requiring that all mass has been lost by the present day.
For the constant radius approximation we obtain:
\begin{equation}
\label{constRdest}
\frac{m^2_i}{R_i^3}= \frac{2}{3}\frac{1}{G \langle a \rangle^2} \eta E^{tot}_X,
\end{equation}
and for the constant density approximation:
\begin{equation}
\label{constdensdest}
\frac{m^2_i}{R^3_i}= \frac{1}{3}\frac{1}{G \langle a \rangle^2} \eta E^{tot}_X,
\end{equation}
using the fact that the density is constant in Eq.~\ref{constdensdest} to write it in terms of the initial mass and radius and adding the subscript $i$ to the radius in Eq.~\ref{constRdest} to illustrate the similarity.
Both destruction lines thus give us a linear cut-off in the $M^2_P/R^3_P$ vs $\langle a \rangle^{-2}$ plane as suggested by \citetalias{davis} with the two differing by a factor of 2 as a result of the different approximations.  We thus find that a population of planets significantly influenced by thermal evaporation should display a linear cut-off in the $M^2_P/R^3_P$ vs $\langle a \rangle^{-2}$ plane.

\begin{figure}
\includegraphics[width=85mm]{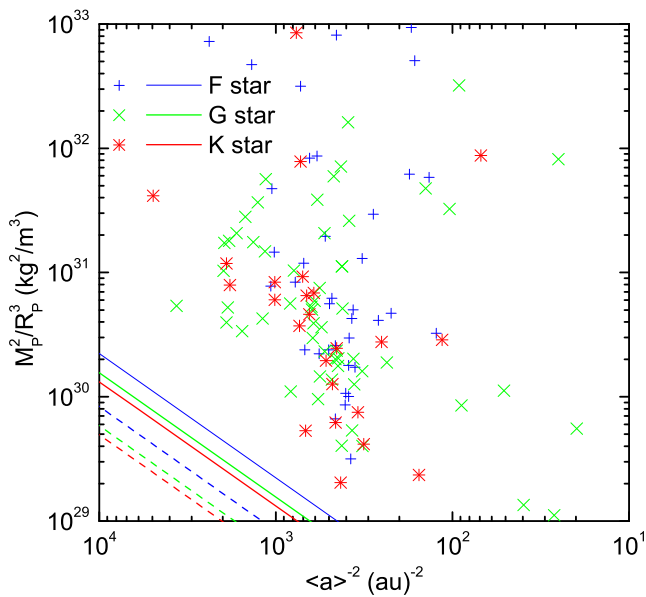}
\caption{Plot of mass squared over radius cubed against the inverse square of the mean orbital distance for our exoplanet sample.  We distinguish planets with F, G and K type hosts and plot the corresponding destruction limits for these spectral types assuming a 4~Gyr age.  Solid lines correspond to the destruction limits for $\eta=1$ under the constant density regime or $\eta=0.5$ under the constant radius regime.  Dotted lines correspond to $\eta=0.25$ for a case intermediate to the constant density and constant radius regimes, or equivalently $\eta=0.375$ for the constant density regime and $\eta=0.188$ for the constant radius regime.  Note that CoRoT-9b and HD80606b lie to the right of the plot area and CoRoT-3b lies to the top of the plot area.}
\label{destplot}
\end{figure}

In Fig.~\ref{destplot} we plot our sample of exoplanets in the $M^2_P/R^3_P$ vs $\langle a \rangle^{-2}$ plane together with destruction limits corresponding to total evaporation of a planet by an age of 4~Gyr under the constant density approximation (assuming that our relations for the unsaturated regime are can be extended to 4~Gyr), i.e. we set $t_0=4$ Gyrs in Eq.~\ref{Extot}.  These destruction limits are not really constraints on the past survival of the known planetary systems.  Rather they test whether analogs of the presently known systems would have survived had they begun their evolution with the masses they have today.  For known systems with young ages the destruction lines place some constraints on the future survival of the system.

Eqs.~\ref{constRdest} and \ref{constdensdest} are linearly dependent on $\eta$ and thus rescaling the destruction lines to different evaporation efficiencies is a simple task.  In Fig.~\ref{destplot} we show two example values, one corresponding to maximal efficiency for the constant density case and one corresponding to a more realistic efficiency of $\eta=0.25$ (as suggested by \citetalias{lammer2009}) for a case intermediate between the constant density and constant radius regimes.  While the upper lines in Fig.~\ref{destplot} correspond to maximal efficiency for the constant density case note that maximal efficiency for the constant radius case would lead to destruction lines a factor of two higher than the upper lines.  Though our formulation for the mass loss will not apply to the three super-Earths 55 Cnc e, CoRoT-7b and Kepler-10b we note that they would fall below, or very close to, both sets of destruction lines in Figs.~\ref{destplot}.

This still does not include all of the processes that will affect planets and their degree of mass loss however.  As noted above the X-ray energy emitted by the host star during the saturated phase will be underestimated by a factor of up to 2, though this may be at least partially counter-acted if, as suggested by \citetalias{lammer2009}, the evaporation efficiency is lower in very strong X-ray irradiation regimes.  \citet*{kashyap} suggest that stars hosting close-orbiting gas-giant planets exhibit an excess of X-ray emission.  These findings apply to older, unsaturated, stars and so would increase the X-ray emission during the unsaturated phase.  We have also not yet discussed the impact of the Roche lobe mass loss enhancement factor, $1/K(\epsilon)$.  These three effects will all tend to push the present population of exoplanets closer to the destruction lines in Figs.~\ref{destplot}.  The ages of some of the planets do of course differ from 4~Gyrs, and we discuss the effects of varying age in Section \ref{ageuncertain}, however 4~Gyrs is the mean age of those hosts with age estimates and we do not expect this to play a large role in the position of the distribution.  Additionally variations in the age of the known exoplanet population does not affect conclusions regarding their long term persistence.

The slope of the destruction lines is quite a good match to the cut-off in the distribution of the known gas-giant exoplanets in the $M^2_P/R^3_P$ vs $\langle a \rangle^{-2}$ plane, which we consider is a required signature of population modification by thermal evaporation.  This slope is an inherent feature of mass loss by thermal evaporation and is independent of most of the other parameters such as the efficiency of the evaporation and the intensity of the stellar X-ray emission.

\subsubsection{Roche lobe effects}
\label{rochedest}

As described above Fig.~\ref{destplot} does not take into account the effects of the proximity of the Roche lobe, which as discussed in Section~\ref{rochelobe} are likely to be quite important.  A useful property of the constant density approximation is that, as we can see from Eq.~\ref{epsilon}, the only planetary parameter on which the Roche lobe mass loss enhancement factor, $1/K(\epsilon)$, depends is the density.  Thus if the density is constant, $K(\epsilon)$ is also constant, making its influence easier to analyse.  Incorporating the Roche lobe factor Eq.~\ref{constdensmi} becomes:
\begin{equation}
\label{constdensmimod}
m_i - m_t = \frac{1}{4 G \pi \rho \langle a \rangle ^2} \frac{\eta}{K(\epsilon)} E_X,
\end{equation}
and the constant density destruction lines becomes:
\begin{equation}
\label{constdensdestKeps}
\frac{m^2_i}{R^3_i}= \frac{1}{3}\frac{1}{G \langle a \rangle^2} \frac{\eta}{K(\epsilon)} E^{tot}_X.
\end{equation}
We can thus think of the effect of introducing the Roche lobe effects as being to change the effective efficiency of the evaporation to $\eta_{eff}=\eta/K(\epsilon)$.

\begin{figure}
\includegraphics[width=85mm]{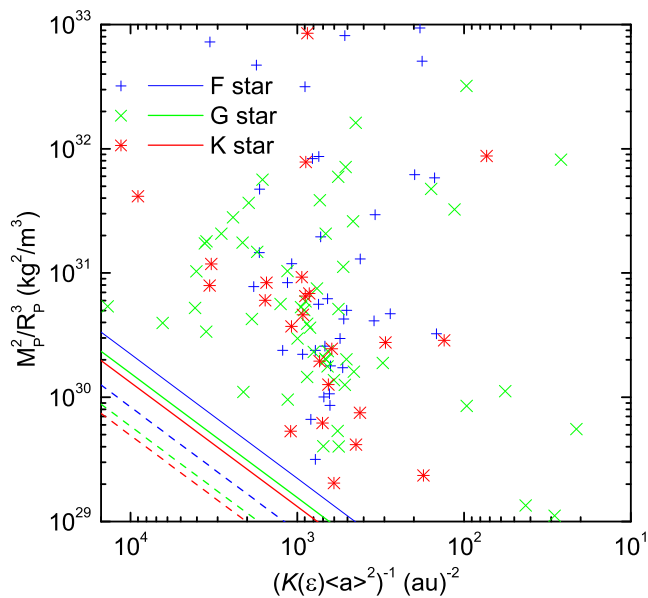}
\caption{After Fig.~\ref{destplot} but now plotting $(K(\epsilon) \langle a \rangle^2)^{-1}$ on the abscissa.  The same destruction lines as on Fig.~\ref{destplot} are shown, but note that this is strictly only relevant to the constant density case, since $K(\epsilon)$ is only constant under the constant density case.  Note that the inclusion of $K(\epsilon)$ adjusts the distribution to lie closer to the destruction lines.}
\label{destplotlow}
\end{figure}

In Fig.~\ref{destplotlow} we now adjust the abscissa of Fig.~\ref{destplot} to now incorporate the Roche lobe factor.  As $1/K(\epsilon)>1$ the effect of this is broadly to move the distribution to the left.  Not all of the planets are shifted equally however since $1/K(\epsilon)$ is different for each of the planets, it is higher for planets with a smaller $\langle a \rangle$ and lower density.  Planets that lie closer to the bottom left in Fig.~\ref{destplot}, and thus which were closer to the destruction lines already, will be shifted furthest.  As a result the distribution is closer to the destruction lines in Fig.~\ref{destplotlow} than in Fig.~\ref{destplot}.  In addition, as we noted at the end of Section~\ref{destlimits} the slope of the destruction lines is independent of the evaporation efficiency and thus a cut-off with this slope is a required signature of a population that has been modified by thermal evaporation.  Importantly the addition of the Roche lobe factor improves the match between the slope of the destruction lines and the cut-off in the distribution.

With the introduction of the Roche lobe factor it is no longer possible to formulate analytic destruction lines for the constant radius case since $1/K(\epsilon)$ will increase over time as the planet loses mass and becomes less dense.  In order to consider the constant radius case on Fig.~\ref{destplotlow} it is necessary to make an approximation about the behaviour of the Roche lobe factor, the simplest of which is to assume that it is constant at the present day value.  For the majority of planets this is quite a good estimate.  We can calculate initial masses for planets in the constant radius regime numerically and thereby capture the true behaviour of the Roche lobe factor (the masses for the constant radius case in Table~\ref{massext} are calculated in this way).  From the true initial masses we can calculate the initial values of $1/K(\epsilon)$ and these are, on average, only 1.5\% lower than the present values with the largest changes being $\sim$20\%.  Thus while the destruction lines in Fig.~\ref{destplotlow} are not strictly correct for the constant radius regime the difference will not be dramatic.

\begin{figure}
\includegraphics[width=85mm]{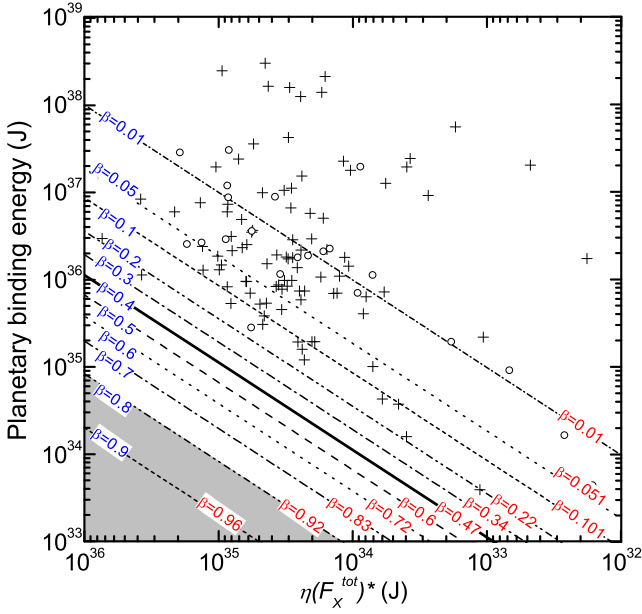}
\caption{Plot of planetary binding energy against $\eta(F^{tot}_X)^*$, a measure of the lifetime X-ray energy absorbed, for the known population of exoplanets assuming $\eta=1$.  For the majority of planets $\eta(F^{tot}_X)^*$ is calculated using the bolometric luminosity of the host star (crosses), where this is not possible the bolometric luminosity of a typical star of the same spectral type is used (circles).  We also plot lines of constant fractional mass loss labelled with the corresponding value of $\beta$ under the constant radius approximation in blue at the left-hand edge and with the values of $\beta$ for the constant density regime in red at the bottom/right.  Planets in the grey-shaded region are at risk of entering a runaway mass loss regime.}
\label{energyplot}
\end{figure}

\subsubsection{Potential energy}
\label{PEroche}

An alternative method to analyse the magnitude of evaporative mass loss is to compare the gravitational potential energy of the planets with the X-ray energy absorbed (as in e.g. \citealt{lecavelier1}; \citetalias{davis}).  As for Eq. \ref{evapequation} to estimate the binding energy of the planets we assume they have the density profile of an $n=1$ polytrope, for which the binding energy is $\frac{3}{4}GM^2/R$.

We also require the time integrated X-ray energy incident on the planet.  We define $(F^{tot}_X)^* \equiv E^{tot}_X \cdot R^2_P/(4\langle a \rangle^2)$, with $R_P$ the radius of the planet today, as a measure of the lifetime X-ray energy incident on the planet.  Under the constant radius approximation this is exactly equal to the time integrated X-ray energy incident on the planet.  Under the constant density approximation the time integrated X-ray energy incident on the planet is not exactly the same as $(F^{tot}_X)^*$ but rather is a function of $(F^{tot}_X)^*$ and the fractional mass loss.

Using this definition for $(F^{tot}_X)^*$ and Eqs.~\ref{constRmi} and \ref{constdensmi} we can, on a plot of binding energy against $\eta(F^{tot}_X)^*$ as a measure of the lifetime X-ray energy absorbed, draw lines of constant fractional mass loss by setting $m_i=m_t/(1-\beta)$ where $\beta$ is the fraction of the initial mass that has been lost.  Using a measure of the lifetime X-ray energy absorbed also allows us to incorporate the (albeit typically rather small) effect of the variation of the ages of the planets in our sample from the mean of 4~Gyrs.  Of the 121 planets in our sample 98 (in 94 separate systems) have age estimates, for the remaining 23 we assume an age of 4~Gyrs, equal to the mean age of the rest of the sample.  Where published $V$ magnitudes, distances and spectral types are available for the host star $(F^{tot}_X)^*$ is determined using the bolometric luminosity calculated for the host.  Some of the host stars lack information on their $V$ magnitudes and/or distances however and in these cases the bolometric luminosity of a typical star of the same spectral type (taken from the tables in \citealt{lang}) is used.  These systems are indicated by an * in Tables \ref{hostext} and \ref{massext}.

\begin{figure}
\includegraphics[width=85mm]{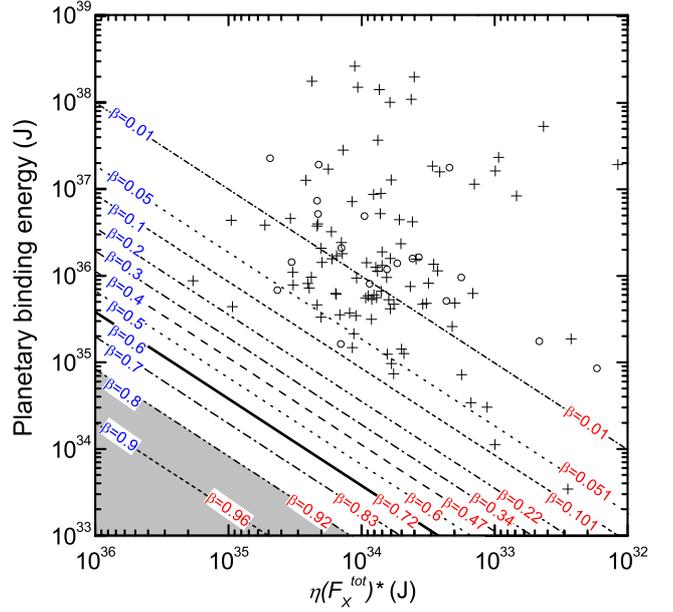}
\caption{As Fig.~\ref{energyplot} but including the Roche lobe reduction to the planetary binding energy and using $\eta=0.25$.  The meaning of the crosses and circles and values of $\beta$ is as before.  With respect to Fig.~\ref{energyplot} the change in $\eta$ induces a horizontal shift in the distribution while the introduction of the Roche lobe factor induces a vertical shift in the distribution, orthogonal to that induced by the change in $\eta$.}
\label{energyplotinc}
\end{figure}

Fig.~\ref{energyplot} shows such a plot of planetary binding energy against $(F^{tot}_X)^*$, in which we use $\eta=1$.  The lines of constant fractional mass loss differ for the constant radius and constant density approximations as a result of the deviation of $(F^{tot}_X)^*$ from the true lifetime X-ray energy incident under the constant density approximation.  The constant density lines take into account the variation of the true lifetime X-ray energy incident on the planet with the fractional mass loss and are exact under the assumption of constant density.  In Fig.~\ref{energyplot} we see that there is a high concentration of planets around the $\beta=0.05/0.051$ (constant radius/constant density) line with a smaller number of systems near the $\beta=0.2/0.22$ line.  We consider that any planet lying in the grey-shaded region would be at risk of entering a runaway mass loss regime.  This would correspond to a $\beta$ of over 0.9 under the constant density approximation and over 0.8 for any reasonable evolutionary path.

In Fig.~\ref{energyplot} we use $\eta=1$ and ignore the Roche lobe correction factor to the binding energy of the planet, $1/K(\epsilon)$.  Adjusting the efficiency will simply shift the entire distribution to the right in Fig.~\ref{energyplot} by a constant factor and has no effect on the lines of constant fractional mass loss.

While the effect of the Roche lobe correction factor, $K(\epsilon)$, is simply to reduce the present planetary binding energy the effect on the mass loss history is more complex.  As we noted in Section~\ref{rochedest} above $K(\epsilon)$ is constant under the constant density regime so in this case the effect of introducing $K(\epsilon)$ is to shift the distribution downwards.  Again there is no change to the lines of constant fractional mass loss, though the downward shift will be different for each planet.  Under the constant radius regime however $1/K(\epsilon)$ will be smaller (closer to 1) at earlier times with the variation dependent on the degree of mass loss.  As described above for most planets this variation will be a small effect and we can provide a reasonable estimate of the impact of the Roche lobe factor by using the present day value.  As the initial value of $1/K(\epsilon)$ will be smaller than the present value this means that the constant radius values of $\beta$ (blue) in Fig.~\ref{energyplotinc} will be overestimates.  For low values of $\beta$ the degree of overestimation will be insignificant, while for larger values of $\beta$ the degree of overestimation will be greater, but still only of the order of a few per cent.

\subsubsection{Effects of age uncertainties}
\label{ageuncertain}
Although as stated above we use the estimated age of the planetary system (where available) to determine the X-ray energy absorbed over the lifetime of the planet and the resulting mass lost since formation these estimated ages are often subject to large uncertainties (typically $\ga50\%$).  Of the 98 planets (94 systems) with age estimates only 11 of these have ages less than 1 Gyr and only 2 less than 0.6 Gyr.  The X-ray luminosity of the host falls by $\sim 2$ orders of magnitude from its peak, saturated, value within the first Gyr and as a result for a system with a typical age of 4~Gyrs the X-ray emission during the first Gyr accounts for $\sim75\%$ of the total lifetime emission.  This means that even though the ages of the exoplanetary systems are subject to rather large uncertainties a change from an age of 4~Gyrs to 1~Gyr will only reduce the fractional mass loss by $\sim25\%$.  Similarly an increase from 4~Gyrs to an age of 10~Gyrs will only increase the fractional mass loss by $\sim15\%$, assuming that our X-ray emission relations can be extended out to 10~Gyrs.  Thus for ages beyond about 1~Gyr the fractional mass loss is comparatively insensitive to the age of the planet.  As a corollary we expect that if a planet is going to be completely stripped of its envelope to leave a chthonian super-Earth this will most likely happen within the first Gyr of its life.

Within the first Gyr of the planet's evolution, and certainly within the first few 100~Myrs (particularly during the saturated period of the host star), the fractional mass loss will vary much more strongly with age.  As such uncertainties in age for those systems with age estimates of $<$1~Gyr will lead to larger variations in the predicted fractional mass loss, though in general the ages of young stars can be estimated more accurately than those of older stars.  Additionally any planet with an estimated age of $\sim$1~Gyr or less that is predicted to have lost a substantial amount of mass must also still be losing mass today at a high rate if the estimated age is correct.  An interesting case is that of WASP-19b, with $\beta=0.06$ for $\eta=0.25$ and an age of only $\sim$1~Gyr.  If this age is correct then we expect that WASP-19b should still be losing mass at a rate substantially higher than that of HD209458b.  If it is in fact older than this, and with a lower present rate of mass loss, then in return the value of $\beta$ must be higher, $0.07$, for an age of 4~Gyrs.

As the study of exoplanets develops and the number of known planets grows it will thus become interesting to look for differences in the mass distribution at different ages.  The magnitude of any differences found would enable constraints to be placed on the value of $\eta$ and thus the impact of evaporation on the evolution of close orbiting exoplanets.

\subsection{Planetary migration}
\label{migration}
The discussion in Section \ref{ageuncertain} also has a bearing on the migration of planets.  The importance of the evaporation that takes place during the earliest stages of the life of the system means that if a planet spends a significant fraction of these early stages at a larger semi-major axis where evaporation is weaker this would substantially alter the total amount of mass lost by the planet.  The key time-scale to compare against here is the saturated period of $\sim$100~Myrs.  In the case of standard disk migration this does not pose any problems since this must be completed on the $\la$6~Myr lifetime of the protoplanetary disk (e.g. \citealt{mamajek2009}).

The Rossiter-McLaughlin effect allows access to the (sky projection of) the angle between the stellar rotation axis and the normal to the planetary orbit for transiting planets, as described by e.g.\ \citet{winn2007}.  Recently it has been discovered through measurements of this effect that a significant number of exoplanets have large misalignments between the planetary orbit and the stellar rotation (e.g.\ \citealt{triaud2010}).  This has prompted suggestions of alternative, non-disk based, methods of placing hot-Jupiters at their present semi-major axes such as planet-planet scattering and migration driven by the Kozai mechanism.  Unlike disk migration these alternative methods have the potential to occur later in the life of the system and thus impact on planetary evaporation.

\begin{figure}
\includegraphics[width=85mm]{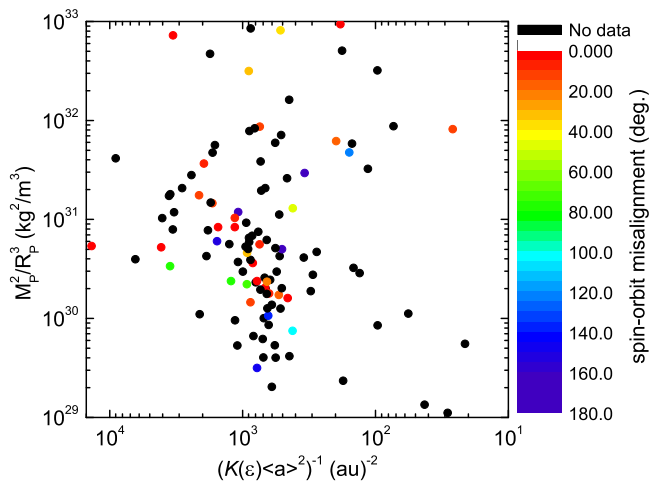}
\caption{Plot of mass squared over radius cubed against the inverse square of the mean orbital distance for our exoplanet sample, including the Roche lobe factor, as in Fig.~\ref{destplotlow}.  Here the misalignment angle between the planetary orbit and the stellar rotation axis indicated by the colour of the points.  Black points indicate planets for which spin-orbit alignment data is not yet available.}
\label{destmisalign}
\end{figure}

Two possibilities arise for evaporation in the light of non-disk based migration.  If the migration happens early then there will be little effect on the evaporation and the constraints on the distribution due to evaporation will be the same as those in the case of disk migration.  On the other hand if non-disk based migration generally happens later, on timescales of the order of 100~Myrs, there would be a significant impact on migration and we would not expect planets that migrated under these mechanisms to be as strongly influenced by evaporation.  If both disk based and non-disk based migration are important with some planets undergoing one and some the other (as suggested by e.g. \citealt{fabrycky2009}) then it would be expected that there would be two distributions in \hbox{$M_P^2/R_P^3$ -- $\langle a \rangle ^{-2}$} space.  The distribution of planets that underwent disk migration would (assuming that evaporation is, as we believe, an important effect) display a linear cut-off in this plane as described in Section \ref{destlimits}.  The distribution of planets that underwent non-disk based migration would be indistinguishable from that of the disk migration planets provided that the non-disk based migration is early.  If however the non-disk migration generally happens later (at around 100~Myrs or more) the non-disk migration planets would be allowed to appear substantially below the cut-off in the distribution of disk migration planets.  An additional potential link between evaporation and migration is the possibility of asymmetric mass loss from the planet inducing orbital migration, as described by \citet{boue2012}.

As a test of this we show in Fig.~\ref{destmisalign} our sample of planets in the \hbox{$M_P^2/R_P^3$ -- $\langle a \rangle ^{-2}$} plane coloured by their spin-orbit misalignment angle.  If misaligned systems represent those which underwent non-disk based migration while aligned systems represent those which underwent disk based migration then the present (albeit limited) sub-sample of planets with measured spin-orbit misalignment angles does not suggest any evidence of a lower cut-off for non-disk migration planets.  This implies that non-disk based migration also happens early in the life of a planetary system, i.e. well within the first 100~Myrs.  As the number of planets with measured spin-orbit misalignment angles increases we will be enabled to better judge whether aligned and misaligned planets obey the same cut-off.

\begin{figure}
\includegraphics[width=85mm]{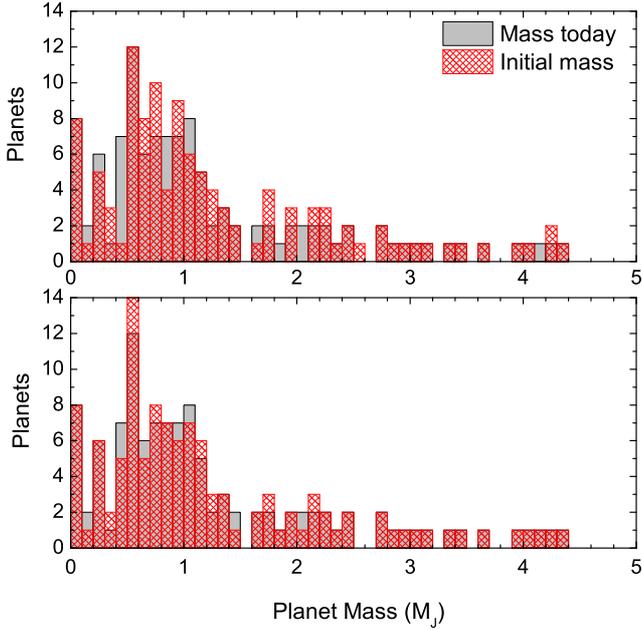}
\caption{Histograms of the masses of our exoplanet sample as observed today, and initial masses as predicted by the energy-limited model under the constant density approximation.  Upper: using $\eta=1$.  Lower: using $\eta=0.25$.  The distributions change very little above 5$M_J$ so we cut off the plots here.}
\label{masshist}
\end{figure}

\begin{table*}
\begin{minipage}{175mm}
\caption{Predictions for mass lost by the planets in our sample.  We list the present mass of the planet ($M_P$) in column (2) and then, for the different regimes discussed in the text, the initial masses in columns (3)--(6) with the fraction of the initial mass lost ($\beta$) to reach the present day mass from the initial mass in columns (7)--(10).  All masses are measured in Jupiter masses $(M_J)$.  Columns (3), (4), (7) and (8) correspond to $\eta=1$ while columns (5), (6), (9) and (10) correspond to $\eta=0.25$.  The constant radius approximation is used for columns (3), (5), (7) and (9) while the constant density approximation is used for columns (4), (6), (8) and (10).  See Sections \ref{PEroche} and \ref{initmasses} for further detail.  The full Table is available online.}
\label{massext}
\begin{center}
\begin{tabular}{l r@{\hspace{25pt}}rrrr@{\hspace{25pt}}rrrr l}
\hline
 & & \multicolumn{4}{c}{initial mass, $m_i$} & \multicolumn{4}{c}{fraction of initial mass lost, $\beta$} &\\
Planet &$M_P$& \multicolumn{2}{c}{$\eta=1$} & \multicolumn{2}{c}{$\eta=0.25$} & \multicolumn{2}{c}{$\eta=1$} & \multicolumn{2}{c}{$\eta=0.25$} & \\
 && cons. R& cons. $\rho$ & cons. R & cons. $\rho$ & cons. R& cons. $\rho$ & cons. R & cons. $\rho$ &\\
(1) &(2)& (3)\hspace{6pt} & (4)\hspace{6pt} & (5)\hspace{6pt} & (6)\hspace{6pt} & (7)\hspace{6pt} & (8)\hspace{6pt} & (9)\hspace{6pt} & (10)\hspace{6pt} &\\
\hline
CoRoT-1 b & 1.030 & 1.4333 & 1.5456 & 1.1491 & 1.1589 & 0.2814 & 0.3336 & 0.1036 & 0.1112 & \\ 
CoRoT-10 b & 2.750 & 2.7503 & 2.7503 & 2.7501 & 2.7501 & 1E-4 & 1E-4 & 2E-5 & 2E-5 & \\ 
CoRoT-11 b & 2.330 & 2.3904 & 2.3917 & 2.3453 & 2.3454 & 0.0253 & 0.0258 & 0.0065 & 0.0066 & \\ 
CoRoT-12 b & 0.917 & 1.0720 & 1.0880 & 0.9586 & 0.9598 & 0.1446 & 0.1572 & 0.0434 & 0.0445 & \\ 
CoRoT-13 b & 1.308 & 1.3287 & 1.3288 & 1.3132 & 1.3132 & 0.0155 & 0.0157 & 0.0040 & 0.0040 & \\ 
\hline
\end{tabular}\\
\end{center}
*$V$ mag.\ or distance information is lacking for the hosts of these planets and thus the predicted fractional mass losses were calculated using the bolometric luminosity of a typical star of the same spectral type as the host (taken from the tables presented in Lang 1991).
\end{minipage}
\end{table*}

\subsection{Predicted initial masses and fractional mass loss of known exoplanets}
\label{initmasses}
Having discussed the effects of evaporation efficiency and the Roche lobe correction to planetary binding energy we now consider the initial masses and fractional mass loss predicted for our sample of known exoplanets.  In Fig.~\ref{masshist} we plot a histogram of the masses today compared with the predicted initial masses for an evaporation efficiency of $\eta=1$ and $\eta=0.25$.  In both cases there are noticeable differences between the present day mass distribution and the predicted initial mass distribution, with the difference being more marked for the $\eta=1$ case as would be expected.  The difference between the present day and initial mass distribution is negligible for higher mass planets ($\ga3 M_J$), which again is as would be expected.  The constant radius approximation, which predicts lower fractional mass loss, produces a slightly smaller shift in the mass distribution but the differences are not dramatic.

\begin{figure}
\includegraphics[width=85mm]{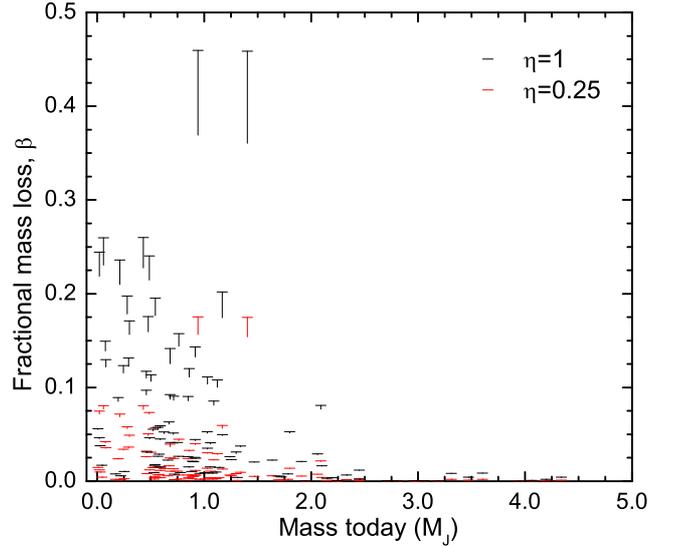} 
\caption{Plot of the predicted fraction of initial mass lost against present day mass for the planets in our sample.  The horizontal dashes indicate the fractional mass loss predicted under the constant density regime, while the bottom of the vertical tail indicates the fractional mass loss predicted under the constant radius regime.  At lower fractional mass loss the difference between the two regimes is negligible such that the vertical tail may not be discernible, while at higher fractional mass loss the regimes diverge.}
\label{fracmassloss} 
\end{figure}

In Table \ref{massext} (full Table available online\footnote{online material web address}) we give for each planet the predicted initial mass and fractional mass loss under both the constant density and constant radius approximation regimes for $\eta=1$ and $\eta=0.25$.  We plot the predicted fraction of initial mass lost against present day mass in Fig.~\ref{fracmassloss}.  Planets near the upper envelope of the distribution represent some of the most closely orbiting examples of planets of their mass and so are, for their mass, some of the easiest to detect given the biases inherent in planet surveys.  As such while additional planets may be discovered in future that lie at higher fractional mass losses we do not expect the distribution to change dramatically, in particular with planets that have present day masses $\ga 3M_J$ not having been subject to high mass loss.  Similarly biases in planet surveys are the reason for an apparent lack of planets with low present day masses and low fractional mass loss, since such planets would be further from their host stars and so would be more difficult to detect.

The planets with the highest fractional mass loss are HAT-P-32b and WASP-12b with both predicted to have lost $\sim$16-18\% of their mass since formation for an efficiency $\eta=0.25$, rising to $\sim$35-45\% for $\eta=1$.  WASP-12b has one of the shortest orbital periods in our sample at only 26 hours while HAT-P-32b is one of the least dense planets in our sample at 148 kg m$^{-3}$.  Having present masses in the region of 1 $M_J$ the $\beta$'s of these planets also translate into very large absolute mass loss with WASP-12b and HAT-P-32b both having lost at least 0.2~$M_J$ since formation.

\subsection{Minimum survival mass}
\label{minmass}
We can use Eq.~\ref{constdensmimod} to estimate the minimum initial mass for a planet of a given initial (and constant) density to survive for 4~Gyrs at a given orbital distance around a star of a given spectral type by setting $m_t=0$ in the same way as we formulated destruction limits in Section \ref{destlimits}.  In general for a planet undergoing total evaporation the density and $K(\epsilon)$ will decrease with time however, and as a result this will underestimate the minimum mass.  In addition when high fractional mass losses are reached there is a possibility of entering a runaway mass loss regime (as discussed earlier), and some fraction of the planet mass may be in the form of a rocky core.  To account for the possibility of runaway mass loss and a remnant core rather than setting $m_t=0$ we choose instead to set $m_t=0.1m_i$ in Eq.~\ref{constdensmimod}.  In this way we obtain the minimum initial mass for survival for 4~Gyrs, $m_S$, as:
\begin{equation}
m_S=\frac{5}{18}\frac{1}{G \pi \rho \langle a \rangle^2}\frac{\eta}{K(\epsilon)}E^{tot}_X
\end{equation}

\begin{figure*}
\begin{minipage}{175mm}
\includegraphics[width=85mm]{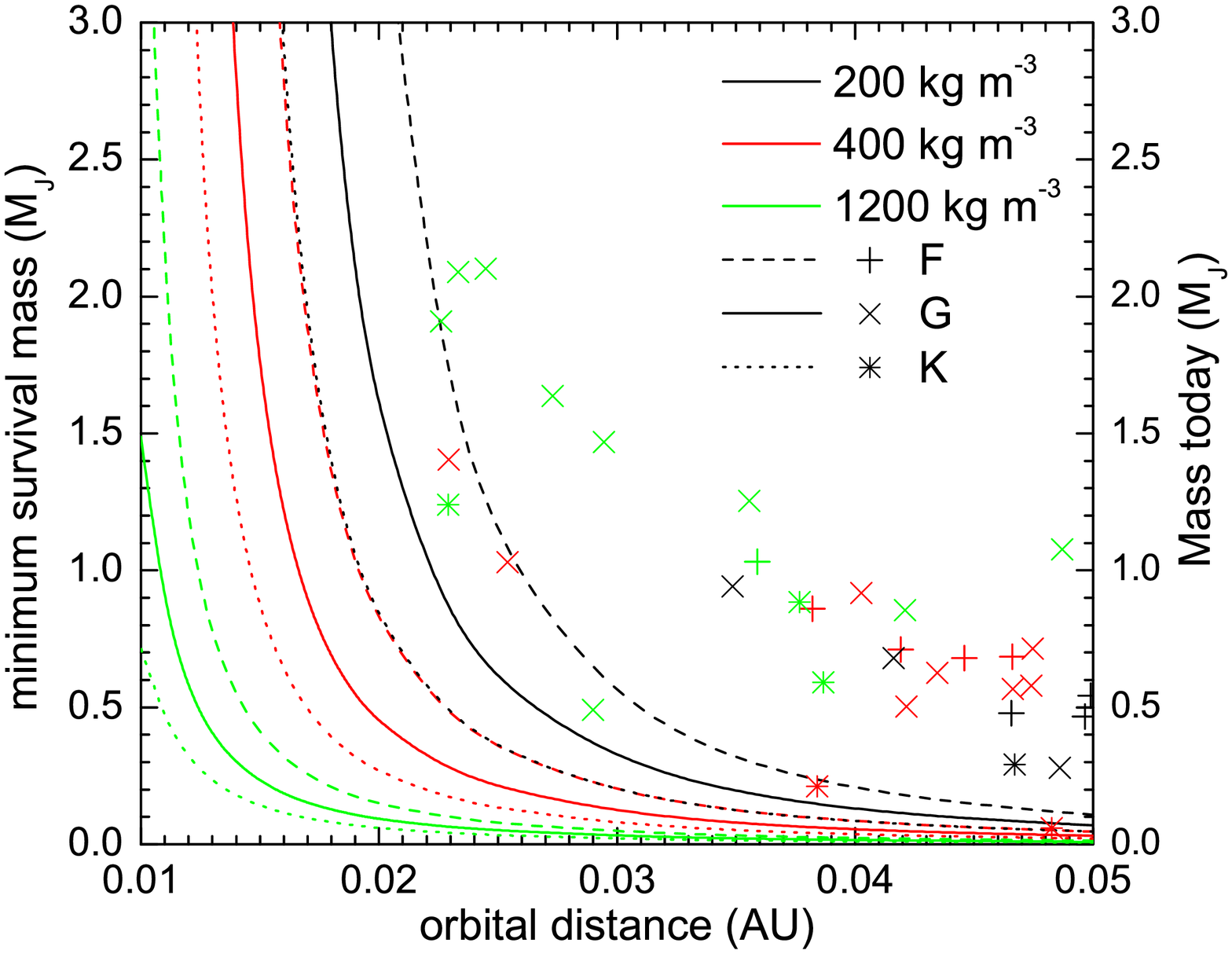}
\includegraphics[width=85mm]{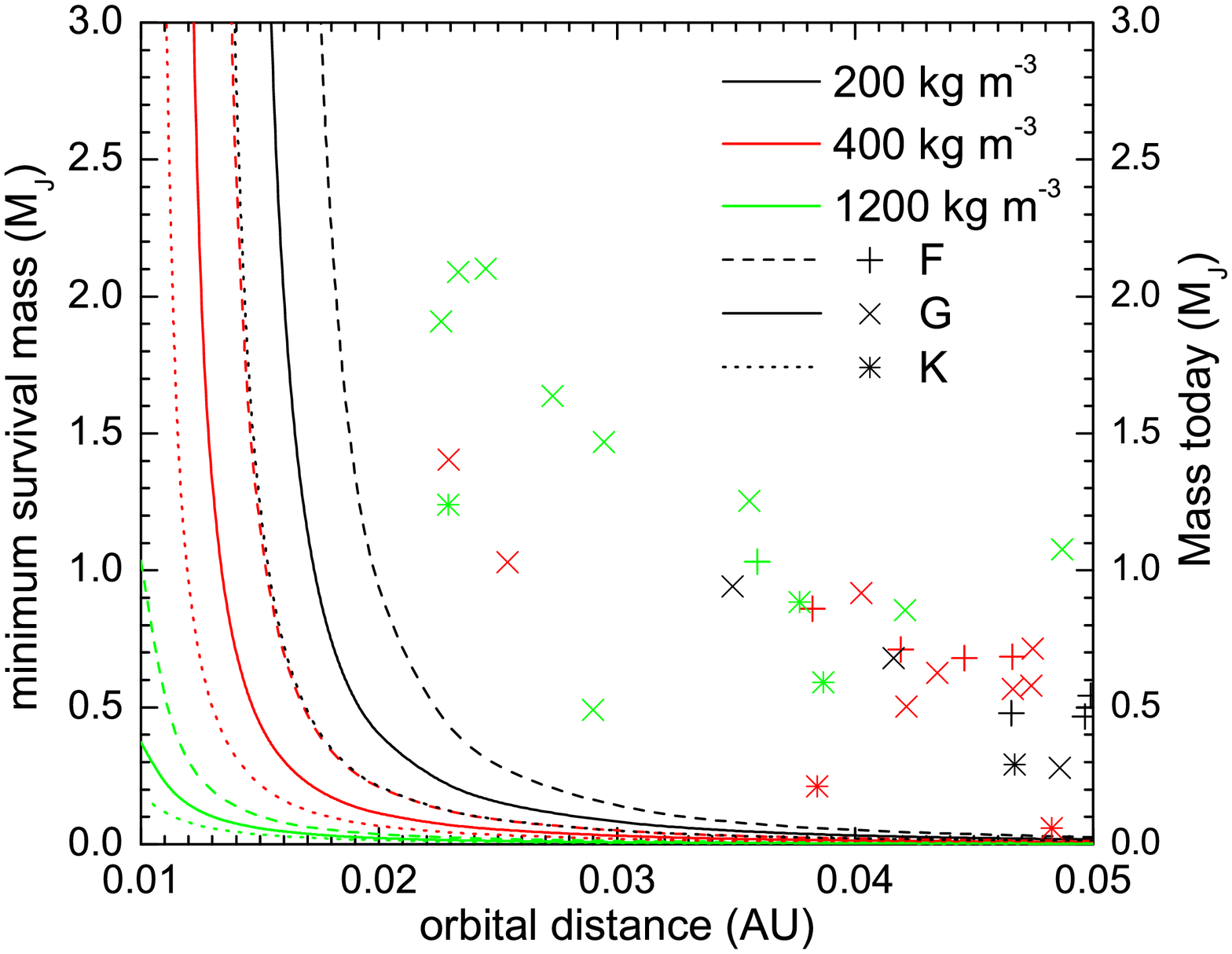}
\caption{Plots of the estimated minimum initial mass for a planet to survive for 4~Gyrs for a selection of different planet densities and host spectral types as a function of orbital distance; left for $\eta=1$, right for $\eta=0.25$.  Included for comparison are the present day parameters of planets from our sample in bands around the densities used for the curves ($<$300~kg~m$^{-3}$, 300-500~kg~m$^{-3}$, 1000-1400~kg~m$^{-3}$).  The colouration of the lines and points indicate the density band to which they correspond while line and point style corresponds to the different spectral types.}
\label{minsurvmasses}
\end{minipage}
\end{figure*}

In Fig.~\ref{minsurvmasses} we present estimates of the minimum initial mass for survival to 4~Gyrs for different spectral types for both $\eta=1$ and $\eta=0.25$ and a selection of representative densities that covers most of the range found for the planets in our sample with masses $\la 2M_J$.  As $m_S$ is linearly dependent on $\eta$ this is easily modified to any desired value of $\eta$.  The highest density used is similar to that of Jupiter (1326~kg~m$^{-3}$) while 400~kg~m$^{-3}$ is roughly the peak of the density distribution for our sample and 200~kg~m$^{-3}$ is typical of the lowest density exoplanets, the lowest density present in our sample being 82~kg~m$^{-3}$ for WASP-17b.

From these figures we can see that planets of the lowest densities found in our sample are essentially precluded from being found at the smallest orbital distances around any spectral type of star.  We also do not expect gas giants of much less than a Jupiter mass to survive at orbital distances of $\la 0.015$AU around F and G type stars unless they are quite dense.  Less massive gas giants might survive at such close orbital distances around K type stars even with more typical densities however.  In comparing the presently known exoplanets with the lines of minimum initial mass for survival we see that for high evaporation efficiencies a number of the presently known exoplanets lie close to or below the survival lines and so we would expect a planet that had been born with the same parameters not to have survived to the present day. As an example for efficiencies $\eta \approx 1$ a planet with an initial mass of $\sim 1 M_J$ and a typical density of $\sim$400~kg~m$^{-3}$ at an orbital distance of $\sim$0.02~AU around an F-type host would today be at most a hot Neptune class planet if not a chthonian super-Earth.  It is also important to bear in mind that even though a planet that is born to a position above, but fairly close to, the relevant line would still have undergone substantial mass loss by an age of 4~Gyrs.  Thus while the planet might 'survive' to at least an age of 4~Gyrs, the 4~Gyr old planet may not bear much resemblance to its younger self.

\subsection{Chthonian planets and the known hot super-Earths}
\label{chthonian}
As we have mentioned earlier in Section \ref{planetsample} the energy-limited evaporation model that we apply to the study of the evaporation of hot-Jupiters is intended for planets with a substantial hydrogen content.  It is unlikely to be suitable for super-Earths of a bulk rock composition as a number of complications become potentially significant such as the latent heat of sublimation and high ionisation states of the large fraction of heavy elements.  Nonetheless a rocky or rock/water composition is not enough to render planets in extremely close orbits such as CoRoT-7b, Kepler-10b and 55 Cnc e immune to evaporation since the stellar irradiation is likely sufficient to melt regions of the crust and even produce a tenuous atmosphere of vapourised silicate minerals (e.g.~\citealt{schaefer2009}, \citealt{leger2011}).  As such the energy limited model can still be used to provide upper limits on the mass loss of super-Earth class planets.  At an evaporation efficiency of $\eta=0.25$ all 3 super-Earths would have lost more than 20 per cent of their mass since formation if they have always had roughly their present compositions.  For 55 Cnc e this would make its initial mass comparable to that of Uranus.  Even for an evaporation efficiency of 0.1 all three would have lost more than half an Earth mass.

This suggests that even a planet that has always had a predominantly rocky composition can be significantly affected by evaporation.  In addition the densities of the super-Earths, being an order of magnitude higher than those typical of close orbiting gas giants, make their mass loss estimates far lower than would be the case for a gas giant under the same conditions.  This raises the question, as also discussed by e.g. \citet{bjackson2010} and \citet{valencia2010}, of whether it is possible that these very close orbiting super-Earths could be chthonian planets -- the remnant cores of gas or ice giants that have been completely stripped of their atmospheres.  Using Fig.~\ref{minsurvmasses} we see that for a K type host a gaseous planet of $\la0.2 M_J$ at the orbital distance of CoRoT-7b or 55~Cnc~e and a typical density of 400~kg~m$^{-3}$ would be completely evaporated by 4 Gyrs.  For a G type host a similar planet of $\la0.3 M_J$ would not survive to an age of 4 Gyrs at the orbital distance of Kepler-10b.  It thus seems possible that these three planets could be chthonian planets though they may also have been born rocky.

\section{Conclusions}
\label{conclusion}
We have used archival X-ray surveys of open clusters to study the coronal X-ray activity-age relationship in late-type stars (in the range $0.29\leq (B-V)_0 < 1.41$), with a particular focus on constraining the regime of saturated X-ray emission.  We find a trend for a decrease in the saturated value of the X-ray to bolometric luminosity ratio across this $(B-V)_0$ range from $10^{-3.15}$ to $10^{-4.28}$ for the latest to earliest type stars in our study.  The saturation regime turn-off ages across the $(B-V)_0$ range of our study are consistent with a scatter around an age of $\sim$100~Myrs.  We also point the reader towards the complementary study of the activity-rotation relation by \citet{wright2011} that was published while this work was in the refereeing process.

In the unsaturated regime we find that the mean value of $\alpha$ in the power law $(L_X/L_{bol})=(L_X/L_{bol})_{sat}(t/\tau_{sat})^{-\alpha}$ is $1.22 \pm 0.10$ with no obvious trend with spectral type.  Under the assumption of an $L_X/L_{bol} \propto \omega^2$ dependence this corresponds to an $\omega \propto t^{-0.61}$ evolution of rotational frequency with time.  The results of our X-ray study are summarised by Eq.~\ref{Lxeq} and Table \ref{satvalues}.  We note that while our unsaturated regime power laws seem broadly consistent with the field star sample of \citetalias{pizzolato} our results should only be extended to stars older than the Hyades with caution since these are the oldest stars in our cluster sample.  Future deep X-ray surveys of old open clusters, or better age estimates for field stars, would enable the evolution of X-ray luminosity in the unsaturated regime to be more strongly constrained.

We have applied our improved constraints on the evolution of the X-ray luminosity of late-type stars to evaporational evolution of close-orbiting exoplanets using the energy-limited model of \citet{lecavelier1} and including a more accurate description of Roche lobe effects as described by \citet{erkaev}.  With a substantially larger sample of planets we confirm the finding of \citetalias{davis} that the planet distribution displays a linear cut-off in the $M_P^2/R_P^3$ vs $\langle a \rangle^{-2}$ plane.  We also confirm that such a cut-off is an expected feature of modification of the population by thermal evaporation irrespective of efficiency for any values typically considered.

We provide estimates of the past thermal mass loss of the known transiting exoplanets, finding that in the case of a constant evaporation efficiency of $\eta=0.25$, 11/121 planets (10 per cent) have lost more than 5 per cent of their mass since formation.  In the case of highly efficient evaporation this fraction rises to a third ($\sim$40/121) of the known transiting exoplanets.  Additionally we calculate estimates of the minimum formation mass for which a planet could be expected to have survived to an age of 4~Gyrs for a range of stellar spectral types, orbital distances, initial planetary densities and evaporation efficiencies.  Both of these suggest that evaporation can have a significant effect on the exoplanet distribution.

It should also be noted that the vast majority of the evaporation occurs within the first Gyr, with the highest evaporation rates during the saturated phase of the host star.  After 1 Gyr we thus do not expect the distribution of planets to change noticeably with the age of the system, however we predict that there should be a significant temporal evolution of the population of planets at ages $<1$ Gyr, with the most marked changes occurring in the first 100-200~Myr, during the saturated phase of the host.  With the increasing rate of exoplanet discoveries and improvements in detection it should soon be possible to test this prediction by comparing planet distributions in young clusters of different ages.

The importance of the earliest phases of evaporation also has implications for planetary migration.  Late migration ($\ga 100$~Myrs) would spare planets from the some of the worst effects of evaporation leading to significantly lower lifetime mass loss.  As work continues to be done on the migration of exoplanets in to close orbits, for example via measurements of the Rossiter-McLaughlin effect, it will be possible to divide the exoplanets into populations that underwent disk migration or non-disk migration.  Differences, or not, between the distributions in the $M_P^2/R_P^3$ vs $\langle a \rangle^{-2}$ plane of planets that arrived at their present orbital positions through different migration mechanisms will allow constraints to be placed on the epoch at which migration occurs.

\section{Acknowledgements}
\label{acknowledgements}
The authors would like to thank the anonymous referee for comments which were helpful in refining this manuscript.  AJ is supported by an STFC Postgraduate Studentship, PW is supported by an STFC rolling grant.  This work has made use of the valuable resources of the SIMBAD database and VizieR catalogue access tool, operated at CDS, Strasbourg, France and the Exoplanet Encyclopaedia, maintained by J. Schneider at the Paris Observatory. The research leading to these results has received funding from the European Community's Seventh Framework Programme (/FP7/2007-2013/) under grant agreement No 229517.

\bibliographystyle{mn2e}
\bibliography{refs}

\label{lastpage}
\end{document}